\documentclass[english,aps,prd,showpacs,preprintnumbers,amsmath,amssymb,onecolumn]{revtex4-1}
\usepackage[T1]{fontenc}
\usepackage[latin9]{inputenc}
\usepackage{amsmath}
\usepackage{amssymb}
\usepackage{graphicx}
\usepackage{hyperref}

\hypersetup{
    bookmarks=true,         
    unicode=false,          
    pdftoolbar=true,        
    pdfmenubar=true,        
    pdffitwindow=false,     
    pdfstartview={FitH},    
    pdftitle={My title},    
    pdfauthor={Author},     
    pdfsubject={Subject},   
    pdfcreator={Creator},   
    pdfproducer={Producer}, 
    pdfkeywords={keyword1} {key2} {key3}, 
    pdfnewwindow=true,      
    colorlinks=false,       
    linkcolor=red,          
    citecolor=green,        
    filecolor=magenta,      
    urlcolor=cyan           
}

\makeatletter

\newenvironment{itemize*}
  {\begin{itemize}
    \setlength{\itemsep}{0pt}
    \setlength{\parskip}{0pt}}
  {\end{itemize}}

\newenvironment{enumerate*}
  {\begin{enumerate}
    \setlength{\itemsep}{0pt}
    \setlength{\parskip}{0pt}}
  {\end{enumerate}}

\newcommand{\GeV}{\,{\rm GeV}}

\newcommand{\beq}{\begin{equation}}
\newcommand{\eeq}{\end{equation}}
\newcommand{\be}{\begin{equation}}
\newcommand{\ee}{\end{equation}}
\newcommand{\bea}{\begin{eqnarray}}
\newcommand{\eea}{\end{eqnarray}}

\newcommand{\nbox}{{\,\lower0.9pt\vbox{\hrule \hbox{\vrule height 0.2 cm
\hskip 0.2 cm \vrule height 0.2 cm}\hrule}\,}}

\def\met {{\not\!\! E_T}}

\@ifundefined{textcolor}{}
{%
 \definecolor{BLACK}{gray}{0}
 \definecolor{WHITE}{gray}{1}
 \definecolor{RED}{rgb}{1,0,0}
 \definecolor{GREEN}{rgb}{0,1,0}
 \definecolor{BLUE}{rgb}{0,0,1}
 \definecolor{CYAN}{cmyk}{1,0,0,0}
 \definecolor{MAGENTA}{cmyk}{0,1,0,0}
 \definecolor{YELLOW}{cmyk}{0,0,1,0}
 }

\def\to{\rightarrow}
\newcommand{\eslash}{\ensuremath{{\hbox{$E_T$\kern-1.2em\lower-.05ex\hbox{/}\kern0.10em\;\;}}}}

\usepackage{amsfonts}
\usepackage{epsfig}
\usepackage{dcolumn}
\usepackage{bm}
\date{\today}

\makeatother
\usepackage{babel}

\begin{document}

\title{Diphotons at the $Z$-pole in Models of the 750 GeV Resonance Decaying to Axion-Like Particles}

\author{Alexandre Alves$^{a}$}%
\email{aalves@unifesp.br}

\author{Alex G. Dias$^{b}$}%
\email{alex.dias@ufabc.edu.br}

\author{Kuver Sinha$^{c}$}
\email{kuver.sinha@gmail.com}

\affiliation{$^a$ Departamento de Ci\^encias Exatas e da Terra, Universidade Federal de S\~ao Paulo, Diadema-SP,
09972-270, Brasil \\
$^b$ Centro de Ci\^encias Naturais e Humanas, Universidade Federal do ABC,
Santo Andr\'e-SP, 09210-580, Brasil \\
$^c$ Department of Physics and Astronomy, University of Utah, Salt Lake City, UT 84112}
\date{\today}

\begin{abstract}
Models in which the 750 GeV resonance ($S$) decays to two light axion-like particles (ALPs $a$), which in turn decay to collimated photons mimicking the observed signal, are motivated by Hidden Valley scenarios and could also provide a mechanism by which a $S \to \gamma \gamma$ signal persists while $S \to Z \gamma,\; ZZ$ and $WW$ remain subdued in the near future. We point out that these Hidden Valley like models invoking $S \to aa \to 4 \gamma$ must also contend with $Z \to a (\to \gamma \gamma) \gamma$ constraints coming from CDF and ATLAS. Within an effective field theory framework, we work out the constraints on the couplings of $S$ to $a$ and gauge bosons coming from photonic $Z$ decays and ensuring that the ALPs decay inside the electromagnetic calorimeter, in two regimes - where $a$ decays primarily to photons, and where $a$ also has hadronic branchings. The analysis is done for both when $S$ has a large as well as a narrow width, and for different relative contributions to the signal coming from $S \to \gamma \gamma$ and $a \to \gamma \gamma$. Results for the particular case where $S$ and $a$ belong to the same complex field are also presented. A $\gamma\gamma$ resonance at the $Z$-pole coming from $Z \to a \gamma$ is expected in this class of models. Taking benchmark ALP masses below around 0.4 GeV and, assuming reasonable values for the fake jet rate and the identification efficiency of the photon-jet, we find the prospects for the discovery of diphotons at the $Z$-pole.
\end{abstract}
\maketitle
\thispagestyle{empty}

\section{Introduction}

One of the most important anomalies in particle physics in recent decades is the excess in $pp\to \gamma\gamma$  peaked at invariant mass around $750 \GeV$ observed at the LHC~\cite{seminar}. We will denote the resonance by $S$ and call it the $S$-cion in this paper, in the hope that $S$ is the first visible scion of a larger dynasty. It is hard to imagine systematic issues, theoretical or experimental, being behind the $S$-cion: the SM background is primarily tree-level $q\bar q\to\gamma\gamma$ scatterings, while experimentally diphotons constitute an extremely clean channel.

CMS presented new data taken without the magnetic field during the Moriond 2016 conference, while ATLAS presented a new analysis with looser photon selection cuts. Moreover, both collaborations recalibrated photon energies optimized around $750$ GeV. The statistical significance of the excess increased for both experiments in the aftermath, leading to renewed activity from theorists. 

For the rates and width of the $S$-cion, we will assume two benchmarks: $(i)$ the narrow width regime with $\sigma_{\gamma\gamma}^{13TeV}=2.5$ fb and $\Gamma_S=5$ GeV; and $(ii)$ the large width regime with $\sigma_{\gamma\gamma}^{13TeV}=6$ fb and $\Gamma_S=40$ GeV. These values follow the fitting of the data presented in ~\cite{Falkowski:2015swt}. 

The literature on the diphoton excess is already vast, covering weakly and strongly coupled models and their embeddings in the UV, as well as new experimental signatures and connections to dark matter and baryogenesis. For a concise summary, we refer to \cite{Strumia:2016wys} and references therein. One particularly interesting class of models that has been proposed is those where the signal arises from {\it photon-jets}. In this class of models, the resonance decays to highly boosted objects which decay to multiple photons. These photons then hit the electromagnetic calorimeter (ECAL) and depending on their angular spread, can be reconstructed as single photons to mimic the signal. Similar proposals have been made previously for the Standard Model Higgs as well, and we refer to \cite{Curtin:2013fra} and references therein for details. Theoretically, such topologies may be motivated by Hidden Valley scenarios \cite{Strassler:2006im}, where the decay $S \to aa$, with $a$ a light scalar or pseudoscalar state, or an axion-like particle (ALP), occurs at tree level. The collimation required to mimic the single photon reconstruction can be obtained if $m_a$ is small, $m_a \sim \mathcal{O}(GeV)$.  

From the perspective of the diphoton anomaly, the main motivation behind this class of models is the ($\sim 1 \sigma$) preference for a wide resonance ($\Gamma/M_S \sim 0.6$) from ATLAS. The reasons are as follows. The simplest realization of weakly coupled models of the $S$-cion (the so-called ``Everybody's Model") consists of loops of vector-like colored and charged matter through which the $S$-cion is produced through $pp$ collisions and subsequently gives the diphoton signal. To realize a wide resonance that simultaneously fits the rates shown above, however, one needs either a large multiplicity of such new particles running in the loop, or large Yukawas or charges, which lead to somewhat baroque models. Moreover, the large coupling of the $S$-cion to gluons required in these scenarios is already constrained by dijet constraints. Models with $gg \to S \to aa \to 4 \gamma$ ameliorate this problem since the coupling to photons occurs at tree-level. 

Another possible motivation to study axion-like models of the new resonance, that is it, models containing the new resonance $S$ and an ALP,  would be a strong suppression of the other electroweak decay channels. If no signals in $Z\gamma$, $ZZ$ and/or $WW$ is observed in the near future, this will force us to consider a different mechanism for the $\gamma\gamma$ decay. In models where the $S$-cion decays to a light scalar or pseudoscalar with large branching ratio to photons, the direct loop-induced decay $S\to\gamma\gamma$, as the other weak bosons channels, can be made small and subdominant. The most recent search for $Z\gamma$ resonances by the CMS Collaboration, by the way, found no signals in the 200-2000 GeV range~\cite{cms-pas-exo-16-021} after combining 19.7 fb$^{-1}$ of the 8 TeV run and 2.7 fb$^{-1}$ of the first 13 TeV run. Also, a combined search for narrow spin-0 and spin-2 resonances in the diboson channels $ZZ$ and $WW$ was performed by the ATLAS Collaboration~\cite{Aaboud:2016okv} with the 3.2 fb$^{-1}$ collected running at $\sqrt{S}=13$ TeV. No excess was found in the 750 GeV.

The purpose of this paper is to point out that Hidden Valley like models invoking $S \to aa \to 4 \gamma$ must also contend with $Z \to a (\to \gamma \gamma) \gamma$ constraints. Just as one expects an excess in $S \to Z \gamma$ and $S \to ZZ$ in the near future simply on the basis of writing down a gauge invariant theory of $S$ that couples to photons, one also expects couplings of $a$ to $Z$ in models where $a$ decays to photons. This opens up $Z \to \gamma \gamma \gamma$ (constrained by ATLAS \cite{Aad:2015bua}) and,  in the limit that $m_a$ is small and the two photons from $a \to \gamma \gamma$ are reconstructed as a single photon, $Z \to \gamma \gamma$ decays (constrained by CDF \cite{Aaltonen:2013mfa}). These constraints are depicted in Fig. \ref{masses}. 

Within an effective field theory framework \cite{Alves:2015jgx}, the coupling of the ALP to $S$ and gauge bosons is constrained from several directions: $(i)$ fitting the diphoton signal; $(ii)$ ensuring that the ALP decays inside ECAL; and $(iii)$ photonic $Z$-decay constraints from ATLAS  \cite{Aad:2015bua} and CDF, as mentioned above. These constraints are presented in two regimes - where $a$ decays primarily to photons and where $a$ also has hadronic branchings. The analysis is done for both the large and narrow width regimes and for different relative contributions to the signal coming from $S \to \gamma \gamma$ and $a \to \gamma \gamma$. Finally, the combined constraints on the space of parameters in the effective field theory are presented. The constraints are also given for the more restrictive case where $S$ and $a$ are the real and imaginary parts of the same complex field.

When the ALP decays exclusively into photons pairs, for example, the CDF data imposes an upper bound on the ALP-photon coupling of 0.07, and with a fifty times stronger bound on  $Z \to \gamma \gamma$ almost all the parameters space of axion-like models of this type can be excluded. These bounds are, of course, weaker if the ALP looks like a genuine axion which decays to gluons but, we show that the constraints from photonic decays of $Z$ bosons need to be taken into account for ALP masses up to $\sim 4$ GeV.

We should expect a $\gamma\gamma$ resonance at the $Z$-pole coming from $Z \to a \gamma$ in this class of models. We take benchmark ALP masses below around 0.4 GeV, where the branching is entirely to photons, and assume what we believe are reasonable values for the fake jet rate and the identification efficiency of the photon-jet. We find that with couplings to the gauge bosons of order 0.07 and with optimal photon detection efficiencies, the LHC will be able to detect diphotons from $Z$ decays with 300 fb$^{-1}$ in a simple cut-and-count experiment. 

The paper is organized as follows. In Section \ref{EFTsection1}, we introduce our notation and the parametrization of the $S$-cion and ALP $a$ in the effective field theory framework. We then discuss, in turn, constraints on our EFT coming from the diphoton signal in Section \ref{diphotonsection1}, the width $\Gamma_S$ of the $S$-cion in Section \ref{widthsection1}, the lifetime of $a$ in Section \ref{Zdecaysection1} and the photonic decays of $Z$ in Section \ref{Zphotonicsection1}. In Section \ref{complexscalar}, we present the constraints in the case where $S$ and $a$ belong to the same complex field. In Section \ref{Zpoleresults}, we present our simulations and results for the search of a resonance at the $Z$-pole. We end with our Conclusions.

\section{Effective Field Theory Parameterization} \label{EFTsection1}

In this Section, we first introduce our notation and the parametrization of the $S$-cion and ALP $a$ in an effective field theory (EFT) framework. We then discuss, in turn, constraints on our EFT coming from the diphoton signal in Section \ref{diphotonsection1}, the width $\Gamma_S$ of the $S$-cion in Section \ref{widthsection1},  the lifetime of $a$ in Section \ref{Zdecaysection1}, and the photonic decays of $Z$ in Section \ref{Zphotonicsection1}. In Section \ref{EFTconstraintsresults}  we put together all the constraints and present our results.

The most general effective Lagrangian involving gauge bosons, the scalar $S$ and the pseudoscalar axion-like $a$ relevant for our studies is given by
\bea
{\cal L} &=& \frac{c_{BB}}{\Lambda}S B^{\mu\nu}B_{\mu\nu}+\frac{c_{WW}}{\Lambda}S W_i^{\mu\nu}W^i_{\mu\nu}+\frac{c_{GG}}{\Lambda}S G_a^{\mu\nu}G^a_{\mu\nu}\nonumber\\
         &+& \frac{k_{BB}}{\Lambda}a B^{\mu\nu}\tilde{B}_{\mu\nu}+\frac{k_{WW}}{\Lambda}a W_i^{\mu\nu}\tilde{W}^i_{\mu\nu}+\frac{k_{GG}}{\Lambda}a G_b^{\mu\nu}\tilde{G}^b_{\mu\nu}\nonumber\\
         &+& \lambda Sa^2 + h.c.
\label{eft}
\eea
where $(\tilde{B}_{\mu\nu},\,\tilde{W}^i_{\mu\nu},\,\tilde{G}^b_{\mu\nu})=\frac{1}{2}\epsilon_{\mu\nu\alpha\beta}(B^{\alpha\beta},\,W_i^{\alpha\beta},\,G_b^{\alpha\beta})$ are the dual field strengths of the $U(1)_Y$ hypercharge, $SU(2)_L$ weak, and $SU(3)_C$ gauge bosons. The effective Lagrangian above can be generated through one-loop diagrams in models having mediators that, beside couplings to $S$ and $a$, interact with the SM gauge bosons as well. If $S$ and $a$ originate from different fields they do not necessarily have the same couplings with the mediators, so that the coefficients $c_{VV}$ and $k_{VV}$, $V=B,\,W,\,G$, in Eq.~(\ref{eft}) do not have to be identical. 

The effective couplings  of $S$ and $a$ with the gauge bosons are each suppressed by their  characteristic energy scales, $v_S$ and $v_a$. These scales can be absorbed in the definitions of the coefficients $c_{VV}$ and $k_{VV}$ through ratios $\Lambda/v_{S,a}$ in favor of the common energy scale $\Lambda$ used in the parametrization of the effective Lagrangian. In fact, $S$ and $a$ may be associated with symmetries that are broken at different energy scales. One example is when the pseudoscalar  $a$ is a pseudo Nambu-Goldstone boson -- and so naturally light -- of a $U(1)_a$ symmetry broken at the scale $v_a$, as happens for the axion or axion-like particles (ALPs), with $S$ being a Higgs boson of another broken symmetry. In Section~\ref{complexscalar}, we will consider the more constrained case where $S$ and $a$ belong to the same complex field and, therefore, have identical couplings to matter implying that $c_{VV}=k_{VV}$. For definiteness, we will in any case call the pseudoscalar $a$ an ALP, considering it as a light pseudoscalar which has a coupling with photons similar to the axion. 
 
The decay of $S \to \gamma \gamma$ is a loop-level process mediated by charged particles, for example, vector-like quarks and leptons. The charged mediators will generally have couplings to the $Z$ bosons through their hypercharge assignments, and with $W$-bosons if they belong to non-singlet representations of $SU(2)_L$. This is also the case of the ALP $a$. In most applications, for example, the detection of axion-like particles, we are interested in the ALP-photon coupling. The ALP-$Z$ coupling thus receives less attention, since light ALPs obviously cannot decay to heavy gauge bosons. Nevertheless, this interaction opens up a new $Z$ boson decay channel, namely, $Z\to a+\gamma$. We refer to \cite{Jaeckel:2010ni} for the status of axion searches with only photon couplings as well as hypercharge couplings, from a combination of Light-Shining-through Wall experiments, cosmology, as well as colliders. 

From the interactions in Eq.~\ref{eft}, we compute the partial widths of $S$ decaying to pairs of ALPs, $a$ to $\gamma\gamma$ and the $Z$ boson decaying to $a+\gamma$
\bea
\Gamma(S\to aa) &=& \frac{\lambda^2}{8\pi m_S}\left(1-4\frac{m_a^2}{m_S^2}\right)^{1/2}\\
\Gamma(S\to \gamma\gamma) &=& \frac{m_S^3}{4\pi\Lambda^2}\times\left(c_{BB}c_w^2+c_{WW}s_w^2\right)^2\\
\Gamma(S\to gg) &=& \frac{m_S^3}{4\pi\Lambda^2}\times 8c_{GG}^2\\
\Gamma(a\to \gamma\gamma) &=& \frac{m_a^3}{4\pi\Lambda^2}\times\left(k_{BB}c_w^2+k_{WW}s_w^2\right)^2\label{Gaaa}\\
\Gamma(a\to gg) &=& \frac{m_a^3}{4\pi\Lambda^2}\times 8k_{GG}^2\label{Gagg}\\
\Gamma(Z\to a\gamma) &=& \frac{m_Z^3}{6\pi\Lambda^2}\left(1-\frac{m_a^2}{m_Z^2}\right)^3\times s_w^2c_w^2(k_{BB}-k_{WW})^2
\eea

The partial decay of the $Z$ boson to three photons is given by
\bea
\Gamma(Z\to \gamma\gamma\gamma) &=& \Gamma(Z\to a\gamma)\times Br(a\to\gamma\gamma)\nonumber \\
&=& \frac{m_Z^3}{6\pi\Lambda^2}\left(1-\frac{m_a^2}{m_Z^2}\right)^3 s_w^2c_w^2(k_{BB}-k_{WW})^2\times Br(a\to\gamma\gamma)
\eea
assuming that $a$ decays predominantly to photons and gluons only. 

\begin{figure}[t]
\includegraphics[width=0.5\textwidth]{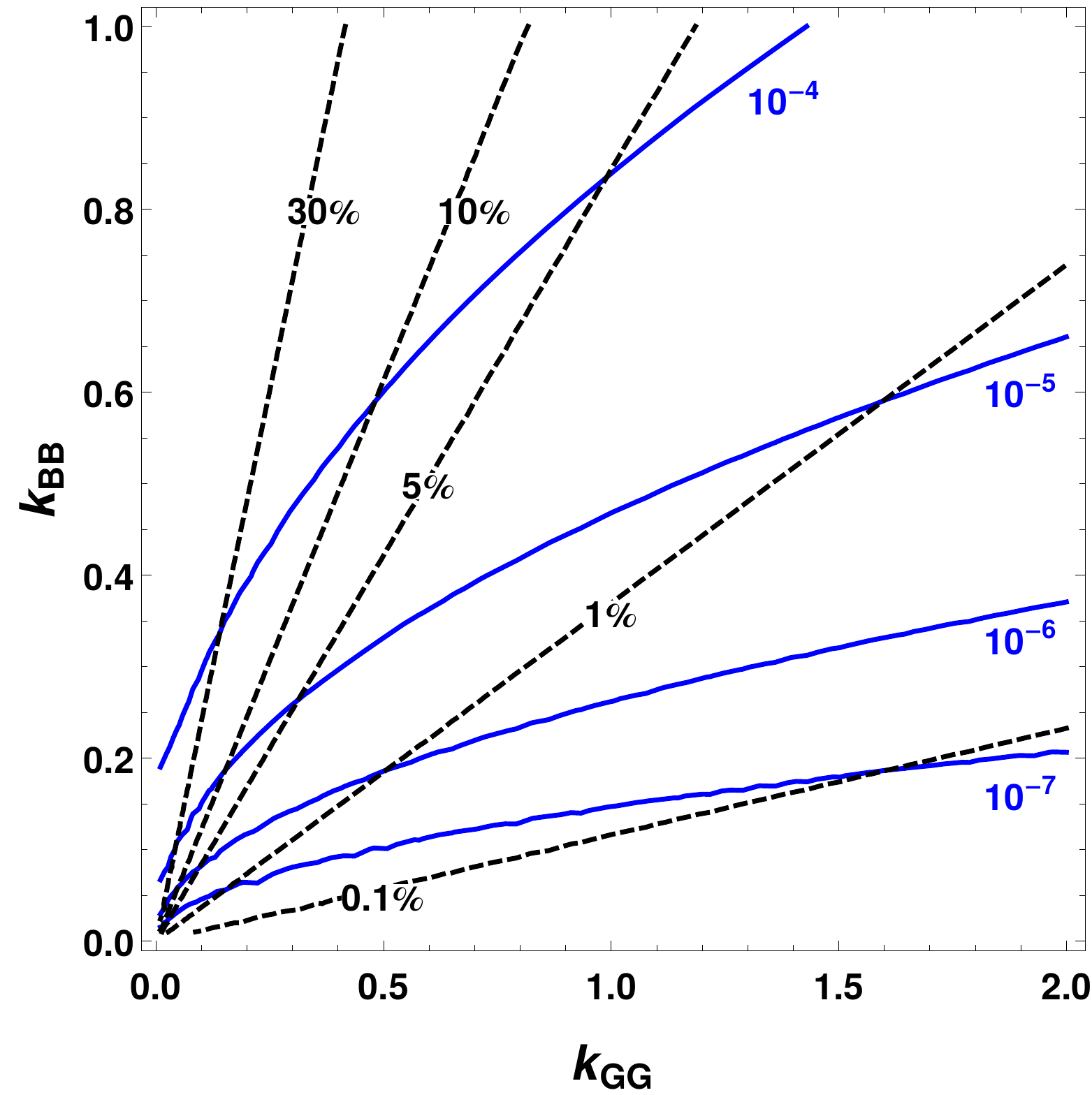}
\caption{Contour lines of constant branching ratios of $Br(Z\to\gamma\gamma\gamma)$, in solid blue, and $Br(a\to\gamma\gamma)$, in dashed black, as functions of the couplings $k_{BB}$ and $k_{GG}$.}
\label{bras}
\end{figure}

Hadronic decays of $a$ are only possible if $m_a>3m_{\pi^0}$, where $m_{\pi^0}\approx 135$ MeV is the neutral pion mass. There are, thus, two regimes we should consider: 
\be
\left\{ \begin{array}{ll}
         Br(a\to\gamma\gamma)=1 & \mbox{if $m_a\lesssim 400$ MeV} \\
         Br(a\to\gamma\gamma)=\frac{1}{1+\frac{8k_{GG}^2}{(k_{BB}c_w^2+k_{WW}s_w^2)^2}} & \mbox{if $m_a > 400$ MeV, including the decay to gluons}
         \end{array} 
\right. 
\label{bra}
\ee
The branching ratios of the ALP into photons pairs and the $Z$ boson into three photons are shown in the Figure~(\ref{bras}).

\subsection{ATLAS and CMS diphoton excess} \label{diphotonsection1}

Having introduced our EFT, we first discuss the constraints on it from the recent diphoton excess. It is useful to parametrize the claimed rates in terms of the following relation involving the partial widths to photons and gluons:
\be
\frac{\Gamma(S\to gg)\Gamma_{\gamma\gamma}}{m_S^2}= C \times\left[\frac{\sigma_{\gamma\gamma}^{13TeV}}{8\hbox{fb}}\right]
\times\left[\frac{\Gamma_S}{m_S}\right]
\label{diphoton}
\ee
where $m_S=750$ GeV and the coefficient $C = 1.1 \times 10^{-6}$. 

While the claimed rates are satisfied if the decay widths to gluons and photons satisfy Eq.~\ref{diphoton}, in our case  $\Gamma_{\gamma\gamma}$ receives two contributions, the first from direct decay $S \to \gamma \gamma$ and the second from  $S\to aa\to\ 4\gamma$. The former is induced by the couplings $c_{BB}$ and $c_{WW}$ of the $S$-cion to the gauge bosons. The latter mimics the diphoton signal and contributes to the claimed rate when the ALP is light. The different mass regimes where the diphoton signal is mimicked are discussed in Section~\ref{Zphotonicsection1} and Fig.~\ref{masses}.

In our analysis, the relative contribution of the channel $S \to aa \to 4 \gamma$ to the total branching to $\gamma \gamma$ is a useful quantity which we denote by
\be \label{Raa}
R_{\gamma\gamma} = \frac{Br(S\to 4\gamma)}{Br_{\gamma\gamma}}=\frac{\Gamma(S\to aa)\times Br^2(a\to\gamma\gamma)}{\Gamma_{\gamma\gamma}}
\ee
where
\be \label{gammagamgamtotal}
\Gamma_{\gamma\gamma} = \Gamma(S\to\gamma\gamma)+\Gamma(S\to aa)\times Br^2(a\to\gamma\gamma) \,\,.
\ee
Using Eq.~\ref{gammagamgamtotal} in Eq.~\ref{diphoton}, we can recast the condition for the claimed rate as
\be
\frac{\Gamma(S\to gg)\left[\Gamma(S\to\gamma\gamma)+\Gamma(S\to aa)\times Br^2(a\to \gamma\gamma)\right]}{m_S^2} = C \times\left[\frac{\sigma_{\gamma\gamma}^{13TeV}}{8\hbox{fb}}\right]\times\left[\frac{\Gamma_S}{m_S}\right] \,\,.
\ee
Finally, using the expressions for $\Gamma(S \to gg)$ and $\Gamma(S \to aa)$ from the EFT Lagrangian in Eq.~\ref{eft} and $R_{\gamma\gamma}$ in the above equation, we arrive at the relation
\be
\boxed{c_{GG}^2\frac{\lambda^2}{\Lambda^2} = \frac{4\pi^2 R_{\gamma\gamma} C}{Br^2(a\to\gamma\gamma)}\times\left[\frac{\sigma_{\gamma\gamma}^{13TeV}}{8\hbox{fb}}\right]\times\left[\frac{\Gamma_S}{m_S}\right]}
\label{digama}
\ee
This provides a relation between parameters of the EFT and the relative contribution of the ``fake photons" that gives the claimed rates for the diphoton excess.

Note that if $Br(a\to\gamma\gamma)=1$, fitting the diphoton signal does not depend on the ALP coupling to the gauge bosons, just $c_{BB},\; c_{GG}$ and $\lambda$. In this case, the constraints that we are going to impose on the model from the photonic decays of the $Z$ boson and the ALP decay length involve just $k_{BB}$ and the mass of the ALP. Whenever we have to take gluonic decays of the ALP into account though, we use Eq.~(\ref{digama}) to eliminate one of those parameters and allowed regions of the parameters space will automatically fit the LHC diphoton signal.

\subsection{Total width} \label{widthsection1}

We now turn to the total width $\Gamma_S$. This is an important piece of information in model building in view of the preliminary results of ATLAS and CMS. In particular, ATLAS data favors a total width of around 40 GeV. On the other hand, CMS data seems not prefer any particular value at this moment and a much narrower resonance is not discarded. As mentioned in the Introduction, we will consider two scenarios inspired by the fitting of the data performed in~\cite{Falkowski:2015swt}: (1) a wide scenario with $\sigma_{\gamma\gamma}^{13TeV}=6$ fb and $\Gamma_S=40$ GeV, (2) a narrow scenario with $\sigma_{\gamma\gamma}^{13TeV}=2.5$ fb and $\Gamma_S=5$ GeV.

\begin{figure}[!t]
\includegraphics[scale=0.5]{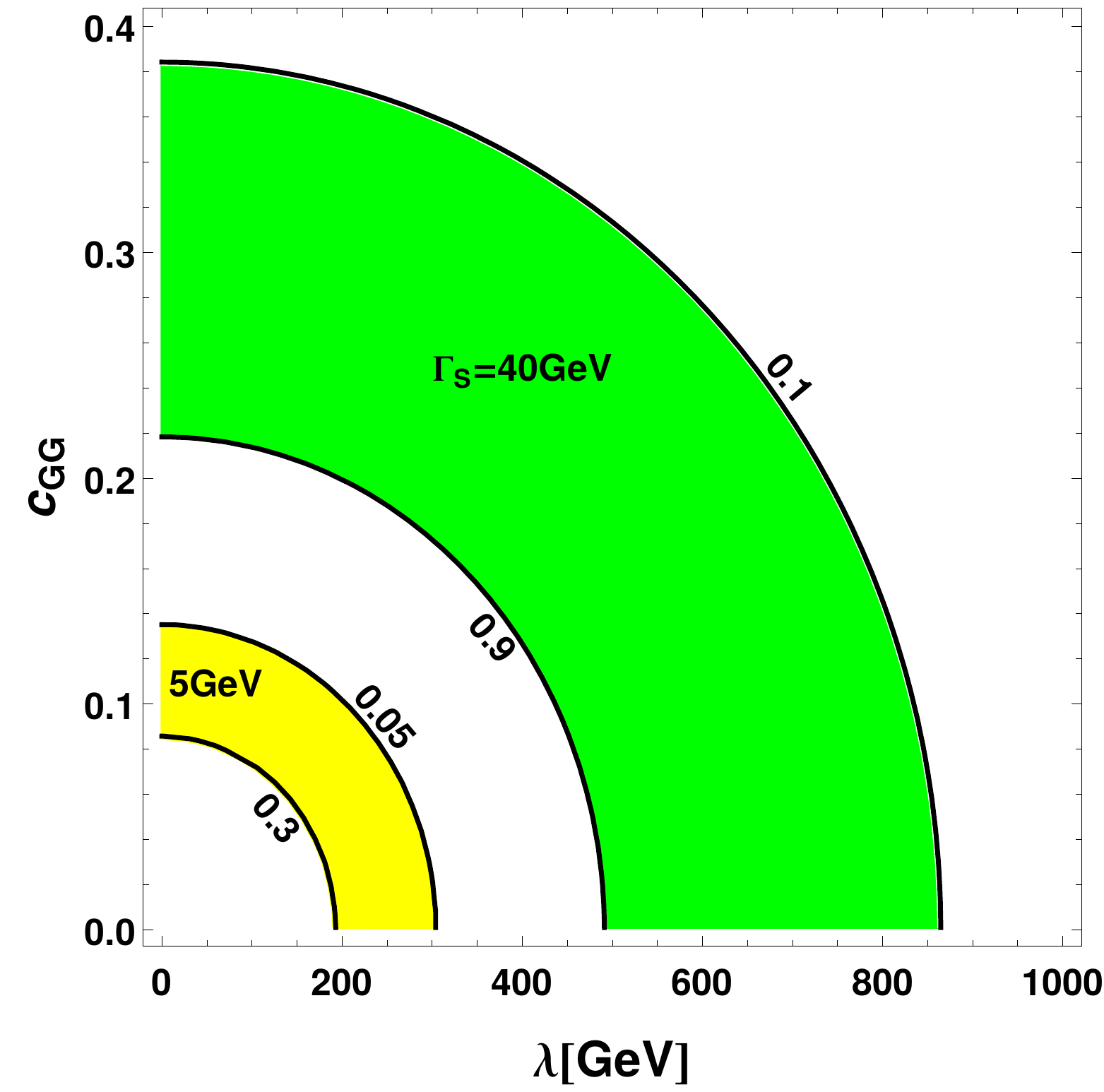}
\caption{The $S$-cion total width of Eq.~(\ref{widthS}) without any dark matter component. The green(yellow) area represents the points with $\Gamma_S=40(5)$ GeV for $c_{BB}$ from 0.1(0.05) to 0.9(0.3).}
\label{width}
\end{figure}

In our EFT parametrization, the total width is given by
\bea
\Gamma &=& \Gamma(S\to aa)+\Gamma(S\to\gamma\gamma)+\Gamma(S\to WW)+\Gamma(S\to ZZ)+\Gamma(S\to Z\gamma)+\Gamma(S\to XX)\nonumber \\
\Rightarrow \Gamma &\approx &\frac{\lambda^2}{8\pi m_S}+\frac{m_S^3}{4\pi\Lambda^2}\left(8c_{GG}^2+c_{BB}^2+c_{WW}^2\right)+\Gamma(S\to XX)
\label{widthS}
\eea
neglecting the ALP mass compared to the $S$ and the $Z$ bosons masses. The partial width $\Gamma(S\to XX)$ represents other contributions we might not be taking into account, for example, the decay into dark matter. Thus, we are allowed to impose the upper bound 
\be
\boxed{\frac{\lambda^2}{8\pi m_S}+\frac{m_S^3}{4\pi\Lambda^2}\left(8c_{GG}^2+c_{BB}^2+c_{WW}^2\right)\leq \Gamma_S}
\label{bound:wid}
\ee
 In the Figure (\ref{width}) we display the total width of Eq.~(\ref{widthS}) without the dark matter contribution $\Gamma(S\to XX)$ in the plane $\lambda$ vs. $c_{GG}$. The green(yellow) area represents the points with $\Gamma_S=40(5)$ GeV for $c_{BB}$ from 0.1(0.05) to 0.9(0.3). As we pointed out, it is easier to get a large width with small $S$-gauge bosons couplings when $S$ decays to ALPs.
%


\subsection{Constraints from ALP lifetime} \label{Zdecaysection1}

The decay width of the ALP $a$ is small, hence it is necessary to ensure that it decays inside the electromagnetic calorimeter. For that propose we have to compute the distance from the interaction point that $a$ can travel before decaying to photons or gluons. In the case of a boosted particle of mass $m_a$ coming from the decay of a heavy particle $S$, this distance is given by
\be
\ell_{decay}=\frac{\beta\gamma}{\Gamma_a}\, [\hbox{GeV}^{-1}]\approx \frac{m_S}{m_a\Gamma_a}\times 10^{-16}\, \hbox{m}
\label{ld}
\ee

In the LHC detectors, the maximum $\ell_{decay}$ is around 1 meter in order that the photons can be detected. In the CDF of Tevatron this distance is not much smaller, around 70 cm~\cite{cdfthesis}. We will consider in both cases a distance of order of a meter to simplify our discussions. We shall see that unless the ALP mass is very small, $\ell_{decay}$ will not represent a severe constraint.

Substituting $\Gamma_a=\Gamma(a\to\gamma\gamma)+\Gamma(a\to gg)$ from Eqs.~(\ref{Gaaa},\ref{Gagg}) into Eq.~(\ref{ld}), we obtain the following bound
\be
\boxed{(k_{BB}c_w^2+k_{WW}s_w^2)^2+8k_{GG}^2 \geq \frac{2\Lambda^2 m_S}{10^{16}\times m_a^4\times \ell_{decay}[\hbox{m}]}}
\label{bound:ld}
\ee
where we are assuming $\ell_{decay}=1$ m.

\subsection{Constraints from photonic decays of the $Z$ boson} \label{Zphotonicsection1}

\begin{figure}[!t]
\includegraphics[scale=0.6]{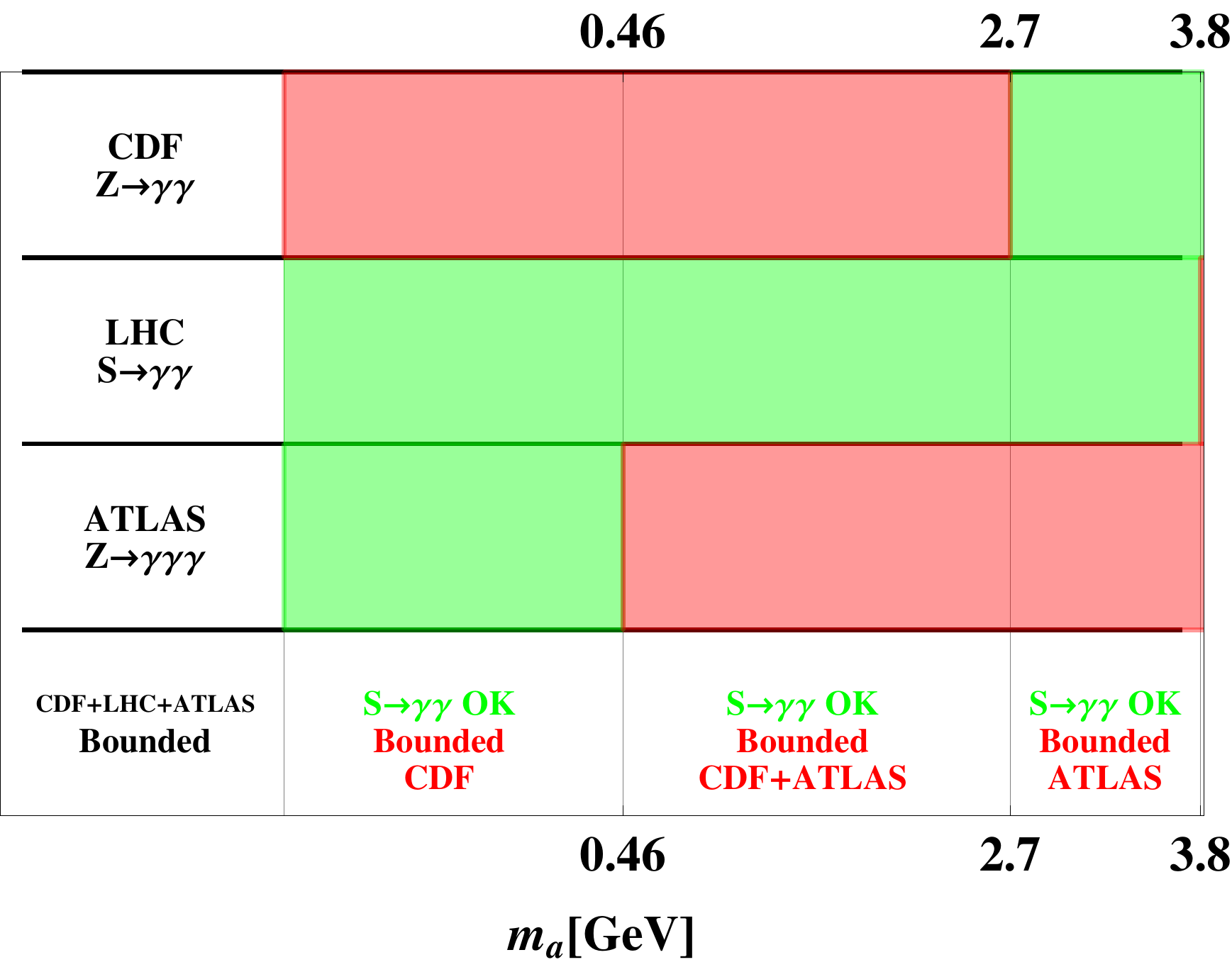}
\caption{In this figure, the red zones represent the ALP masses for which the constraints from searches at for photonic decays of the $Z$ boson at the Tevatron and the LHC apply. The green zones represent masses where either $S$ might decay to two photon-jets that fakes a diphoton signal or no bounds from photonic $Z$ bosons are involved.}
\label{masses}
\end{figure}

In this Section, we turn to the final important constraint: the upper limit from the CDF Collaboration~\cite{Aaltonen:2013mfa} on the $Z$ boson decay to two photons, and from the ATLAS Collaboration ~\cite{Aad:2015bua} on the $Z$ boson decay to three photons. The limits at 95\% of confidence level (CL) are as follows:
\bea
\textrm{CDF} \,\,\,Br(Z\to \gamma\gamma): \,\,\,\,\, 1.45\times 10^{-5} \nonumber \\
\textrm{ATLAS} \,\,\,Br(Z\to 3 \gamma):   \,\,\,\,\, 2.2\times 10^{-6} \,\,.
\eea
The realm of validity of each constraint is depicted in Figure~(\ref{masses}), which we now discuss in detail. Let us point out that, in the Standard Model, $Z\to\gamma\gamma\gamma$ is extremely rare, with a branching ratio of order 10$^{-10}$~\cite{zgammas}.

We start with large ALP masses. Ref.~\cite{Jaeckel:2015jla} studied $Z$ bosons decaying to an axion-like pseudoscalar and a photon. They found that the $Z\to 3 \gamma$ channel can probe ALPs with masses between 4 and 60 GeV where the two photons from $a \to \gamma \gamma$ are more easily resolved. This is therefore the region where the  750 GeV scalar would give rise to a signal with four photons instead of two, and is hence less interesting for us.

As the ALP mass is reduced, the photons coming from $a \to \gamma \gamma$ get more and more collimated. The ALP mass thresholds where the final state photons start to mimic the diphoton signal depend on the mass of the mother particle and the resolution of the detector. The azimuthal opening angle between the two photon jets coming from an initial state $Y \to a \to \gamma \gamma$ is given by
\be \label{azimuthal}
\Delta\phi\sim \frac{4m_a}{m_X} \,\,.
\ee
The angular resolution of the LHC detectors is $\Delta\phi ~ \sim ~ 20$ mrad~\cite{lhc_resolution}. We then obtain, using Eq.~\ref{azimuthal} and putting $m_X = m_Z$, that the ATLAS limits on $Br(Z\to 3 \gamma)$ apply for $m_a \, > \, 0.46$ GeV. In the regime $0.46$ GeV $< \, m_a \,$ $< \, 4$ GeV, then, the photons coming from $S \to aa \to 4 \gamma$ mimic the diphoton signal, but the photons coming from $Z \to a \gamma \to 3 \gamma$ are subject to ATLAS bounds. For ALPs lighter than 0.46 GeV, the photon-jets from $a \to \gamma \gamma$ are too collimated to be resolved at the LHC and the  ATLAS limits on $Br(Z\to 3 \gamma)$ no longer apply. 

We turn now to the regime where CDF constraints on $Z \to \gamma \gamma$ apply. Taking the resolution of the CDF detector to be 120 mrad~\cite{cdfthesis}, we obtain $m_a \, > \, 2.7$ GeV using Eq.~\ref{azimuthal}. We show in Figure~(\ref{masses}) the ALP mass regions relevant for each constraint that we have just discussed.

We now consider the constraints on the EFT parameters coming from photonic decays of $Z$. The branching ratio of the new $Z$ decay channel is
\bea
Br(Z\to 3\gamma)=\frac{\Gamma(Z\to a+\gamma)\times Br(a\to \gamma\gamma)}{\Gamma_Z} &=& \frac{m_Z^3}{6\pi\Lambda^2\Gamma_Z}\left(1-\frac{m_a^2}{m_Z^2}\right)^3 \times \nonumber\\
& & \frac{s_w^2c_w^2(k_{BB}-k_{WW})^2}{1+8k_{GG}^2/(k_{BB}c_w^2+k_{WW}s_w^2)^2}
\eea

Neglecting the ALP mass in comparison to the $Z$ boson we now have another constraint on the parameters in the EFT Lagrangian
\be
\boxed{\frac{(k_{BB}-k_{WW})^2}{1+8k_{GG}^2/(k_{BB}c_w^2+k_{WW}s_w^2)^2}\leq \frac{6\pi^2\Lambda^2\Gamma_Z}{m_Z^3s_w^2c_w^2}\times Br_Z^{exp}}
\label{bound:brz}
\ee
The decay $Z\to \gamma+a$ gives rise to three photons signals, but for a light $a$ a very boosted and collimated photon-jet also emerges from the $a$ decay mimicking a diphoton signal. Thus, we use $Br_Z^{exp}=1.45\times 10^{-5}$ in the case where the CDF limit applies and $2.2\times 10^{-6}$ where the ATLAS limit is relevant from Fig.~\ref{masses}.


\section{Results: Constraining the EFT} \label{EFTconstraintsresults}
In this Section, we put together all the constraints discussed in Sections \ref{diphotonsection1}, \ref{widthsection1}, \ref{Zphotonicsection1}, and \ref{Zdecaysection1} for study the parameter space defined by our EFT. We begin with some simplifications, to render the constraints amenable to a clear exposition. Moreover, all the points allowed in the forthcoming results were checked against constraints from null signals in the $ZZ$, $Z\gamma$, $WW$ and $gg$ channels~\cite{Strumia:2016wys}.  The $Z\gamma$ channel, in particular, was constrained with the recent analysis of the CMS Collaboration~\cite{cms-pas-exo-16-021} from which we infer that $Br(S\to Z\gamma)/Br(S\to \gamma\gamma)\lesssim 2$--$3$ at 95\% CL. Of course, the ALP contribution to the diphoton signal makes it easier to respect those bounds as we can always adjust the ALP-$S$-cion coupling strength $\lambda$.

The general EFT parametrization of Eq.~(\ref{eft}) involves nine parameters: three couplings (the $k$'s) of the ALP $a$ to the gauge bosons, three couplings (the $c$'s) of the scalar $S$ to the gauge bosons, the mass dimensional coupling of the scalars $\lambda$, the ALP mass, and the new physics scale $\Lambda$. 

This number can be reduced to seven assuming that both $S$ and $a$ are $SU(2)_L$ singlets, $c_{WW}=k_{WW}=0$. From now we consider just the singlet models to perform our analysis. This is a well motivated simplification that will help us to illustrate how the constraints that we are considering are important in models with axion-like particles.

We also fix $\Lambda=1$ TeV as this parameters always appears in ratios with the various couplings to the gauge bosons. Now we have six parameters: $k_{GG},\; k_{BB},\; c_{GG},\; R_{\gamma\gamma},\; \lambda,\; \hbox{and}\; m_a$, where we have eliminated $c_{BB}$ in favor of $R_{\gamma\gamma}$ defined in Eq.~(\ref{Raa}).

We can simplify Eq.~(\ref{digama}) for the claimed diphoton rate by assuming  $k_{WW}=0$, obtaining
\be
\frac{c_{GG}\frac{\lambda}{\Lambda}}{1+\frac{8}{c_w^4}\left(\frac{k_{GG}}{k_{BB}}\right)^2}=\sqrt{4\pi^2 R_{\gamma\gamma}C\times\left[\frac{\sigma_{\gamma\gamma}^{13TeV}}{8\hbox{fb}}\right]\times\left[\frac{\Gamma_S}{m_S}\right]}
\label{param1}
\ee

Let us now impose the constraints that we have discussed so far on the axion-like models of the 750 GeV resonance in the EFT approach. We need to consider two regimes.

\subsection{$m_a<3m_{\pi^0}$ ($k_{GG} = 0$)}


 If $m_a$ is less than three neutral pion masses then the ALP decays exclusively to photons and $Br(a\to\gamma\gamma)=1$ as we have already discussed. We can thus take $k_{GG} = 0$ in Eq.~\ref{param1}. This enables us to plot  $c_{GG}$ as a function of the ALP-$S$-cion coupling $\lambda$.

\begin{figure}[!t]
\includegraphics[scale=0.6]{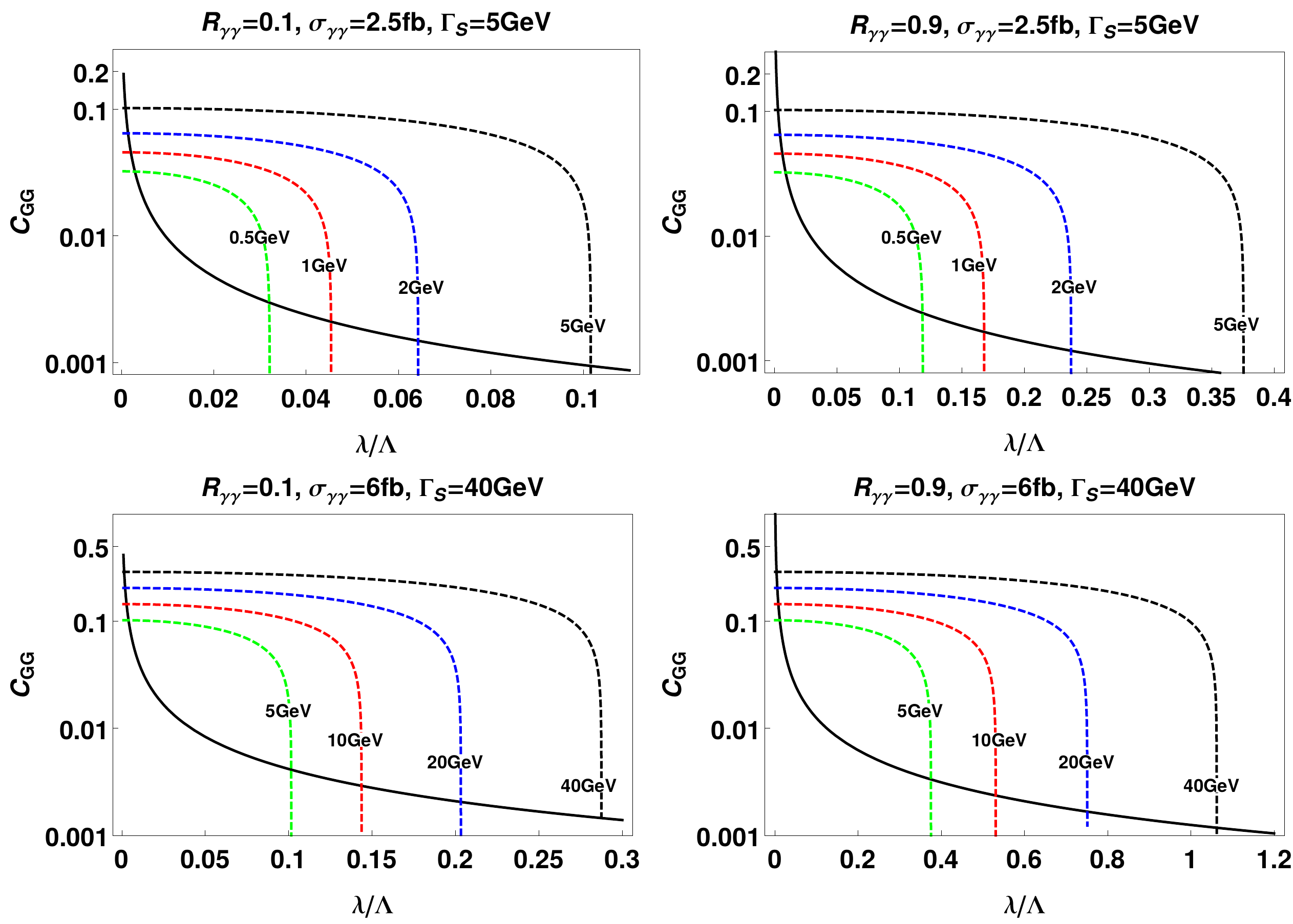}
\caption{$m_a<3m_{\pi^0}$, $c_{GG}$ vs. $\frac{\lambda}{\Lambda}$ plane: The solid black lines show the points on the $c_{GG}$ vs. $\frac{\lambda}{\Lambda}$ space which fit the diphoton excess for $\Gamma_S = 5$ GeV, $\sigma=2.5$ fb (upper two panels) and $\Gamma_S = 40$ GeV, $\sigma=6$ fb (lower two panels), with $R_{\gamma\gamma}$ fixed at 0.1 (left panels) and 0.9 (right panels). We have assumed $m_a<0.4$ GeV. The dashed lines are contours of the width $\Gamma_S$.}
\label{xsecs}
\end{figure}

In Figure~(\ref{xsecs}), the solid black lines show the points on the $c_{GG}$ vs. $\frac{\lambda}{\Lambda}$ plane which fit the diphoton excess for $\Gamma_S = 5$ GeV (upper two panels) and $\Gamma_S = 40$ GeV (lower two panels), with $R_{\gamma\gamma}$ fixed at 0.1 (left panels) and 0.9 (right panels). We have assumed $m_a<0.4$ GeV and used Eq.~\ref{param1}. Further, using Eq.~\ref{bound:wid}, we plot contours of the width $\Gamma_S$ on the same plane, represented by the dashed lines. The intersections of the black solid line with the colored dashed lines shows that there are solution points for a wide range of widths for each scenario. It is implicit in these plots that a large portion of the parameters space for $k_{BB}$ and $m_{a}$ can fit the diphoton signal.

\begin{figure}[!t]
\centering
\includegraphics[scale=0.7]{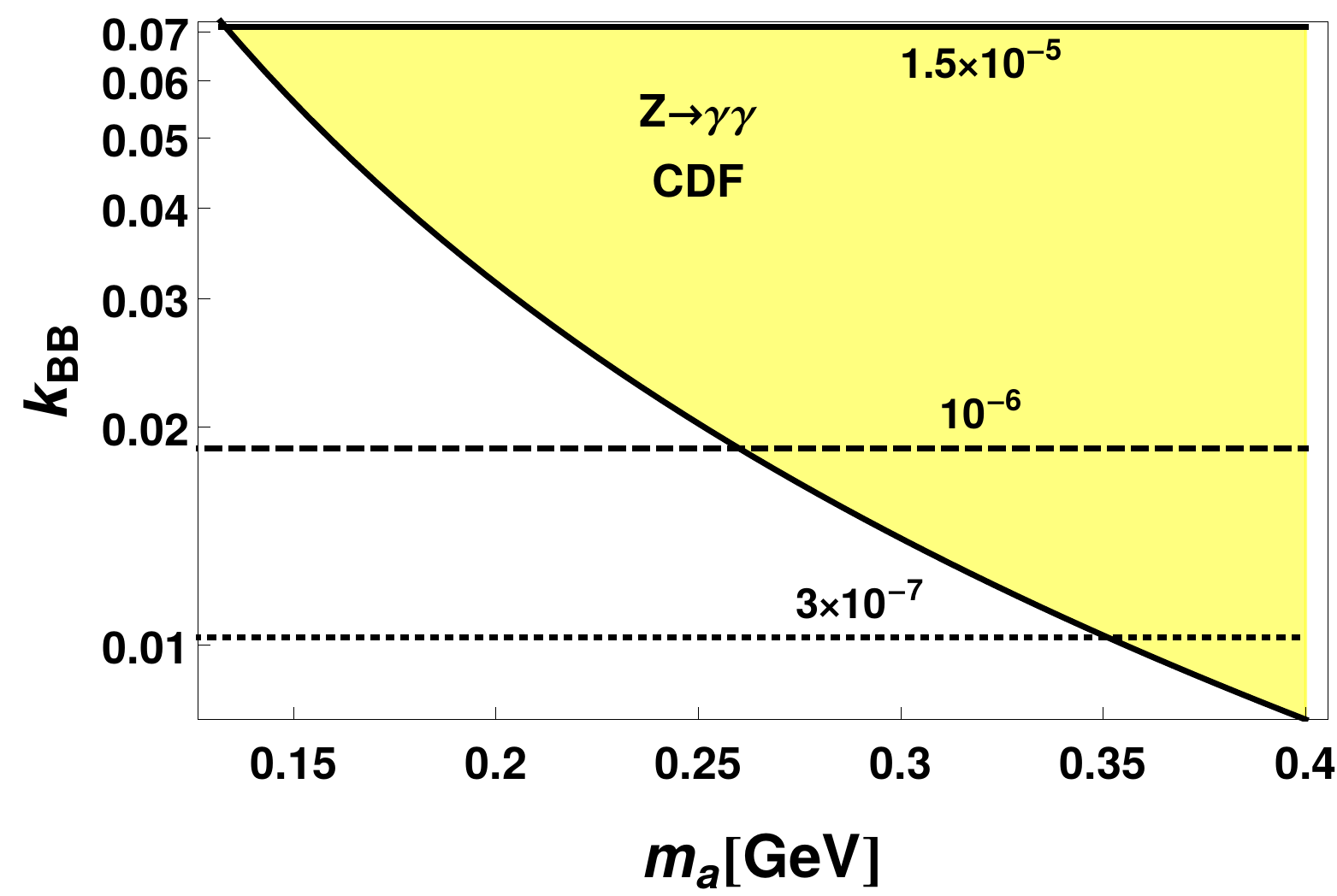}
\caption{$m_a<3m_{\pi^0}$, $k_{BB}$ vs. $m_a$ plane: The yellow allowed region is bounded by the straight line at $k_{BB}=0.07$ and the black solid curve. The region $k_{BB}>0.07$ is excluded from the CDF experimental constraint in this mass region, after using the CDF experimental value in Eq.~\ref{bound:brz}. For a given $m_a$ there is a lower limit coming from Eq.~\ref{bound:ld}, shown by the black solid curve, if we demand that the ALP decays inside the calorimeter of the CDF detector. ALPs lighter than approximately  100 MeV cannot be bounded by the results of these experiments as they decay outside the region of the electromagnetic calorimeter. The dashed lines represent the constraints from $Z\to\gamma\gamma$ for 15 and 50 times stronger bounds from possible future searches. In this plot, $k_{WW} = k_{GG} = 0$.}
\label{fig1}
\end{figure}

We now turn to constraints on $k_{BB}$, which come from the photonic decays of $Z$. From Eq.~\ref{bound:ld} and Eq.~\ref{bound:brz}, and setting $k_{WW} = k_{GG} = 0$, we immediately get upper and lower limits on $k_{BB}$ independently of the other parameters. For $m_a<3m_{\pi^0}\approx 0.4$ GeV, the relevant limit is from the CDF search for photon pairs decays of the $Z$ boson. In Figure~(\ref{fig1}) we show the allowed region of the $k_{BB}$ {\it versus} $m_a$ plane for $m_a<0.4$ GeV. The yellow allowed region is bounded by the straight line at $k_{BB}=0.07$ and the black solid curve. The region $k_{BB}>0.07$ is excluded from the CDF experimental constraint in this mass region, after using the CDF experimental value in Eq.~\ref{bound:brz}. For a given $m_a$ there is a lower limit coming from Eq.~\ref{bound:ld}, shown by the black solid curve, if we demand that the ALP decays inside the calorimeter of the CDF detector. ALPs lighter than approximately  100 MeV cannot be bounded by the results of these experiments as they decay outside the region of the electromagnetic calorimeter. The dashed lines represent the constraints from $Z\to\gamma\gamma$ for 15 and 50 times stronger bounds from possible future searches. If the LHC finds no diphoton signal at the $Z$-pole and the CDF limit can be made two orders of magnitude smaller than the current bound, models with ALPs mimicking the diphoton signal will be strongly disfavored if $m_a<3m_{\pi^0}$. 

Of course, it is possible to invoke new decay channels of the ALP into charged leptons and neutrinos. Investigating these cases could also be very interesting but hard. The irreducible SM background for $Z\to \ell^+\ell^-\gamma$, where the photon comes from the bremsstrahlung emission of the charged lepton, is expected to be much larger than the signal $Z\to a\gamma\to \ell^+\ell^-\gamma$, and in both cases, the final state reconstructs the $Z$ boson. Decays to neutrinos would lead to monophoton signals which have been scrutinized by the experiments but, in this case, the photon transverse momentum is not too hard, around half $m_Z$. Actually, as there are no bounds for $Z\to\gamma+\met$, this channel could work if the branching ratio is large.

\subsection{$m_a\geq 3m_{\pi^0}$ ($k_{GG} \neq 0$)}

When the ALP is allowed to decay into gluons we have another parameter coming into play: $k_{GG}$. However, as in the previous case, the constraints from $Z$ decays to photons inside the calorimeter region do not depend on $c_{GG},\; c_{BB}$ and $\lambda$, just on the couplings of the ALP with the gauge bosons and the ALP mass. On the other hand, once we have chosen an allowed point we can compute $Br(a\to \gamma\gamma)$ from Eq.~(\ref{bra}) and the product $c_{GG}\frac{\lambda}{\Lambda}$ which fits the diphoton excess from Eq.~(\ref{diphoton}) for a given $R_{\gamma\gamma}$.

\begin{figure}[!t]
\includegraphics[scale=0.5]{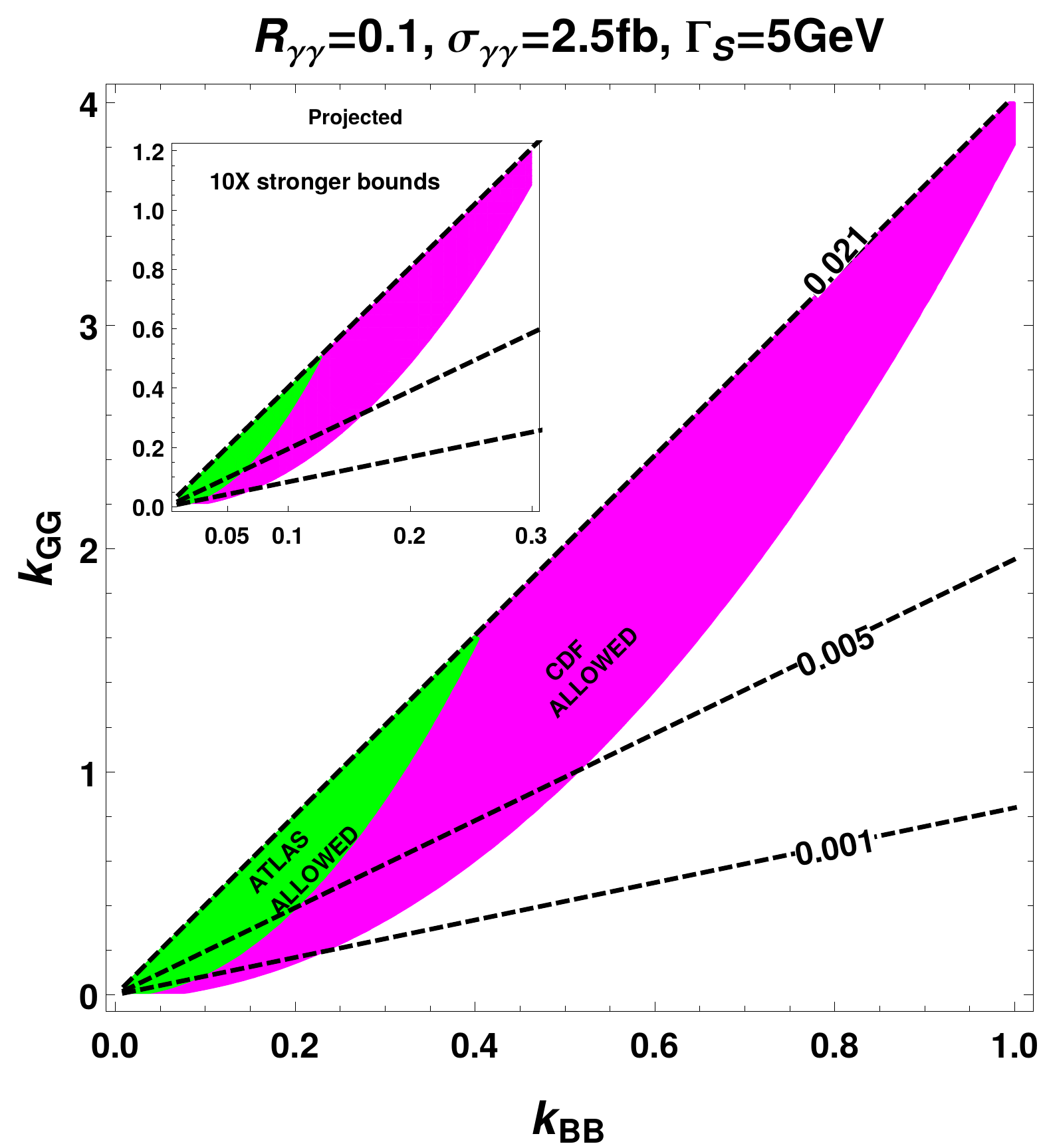}
\includegraphics[scale=0.5]{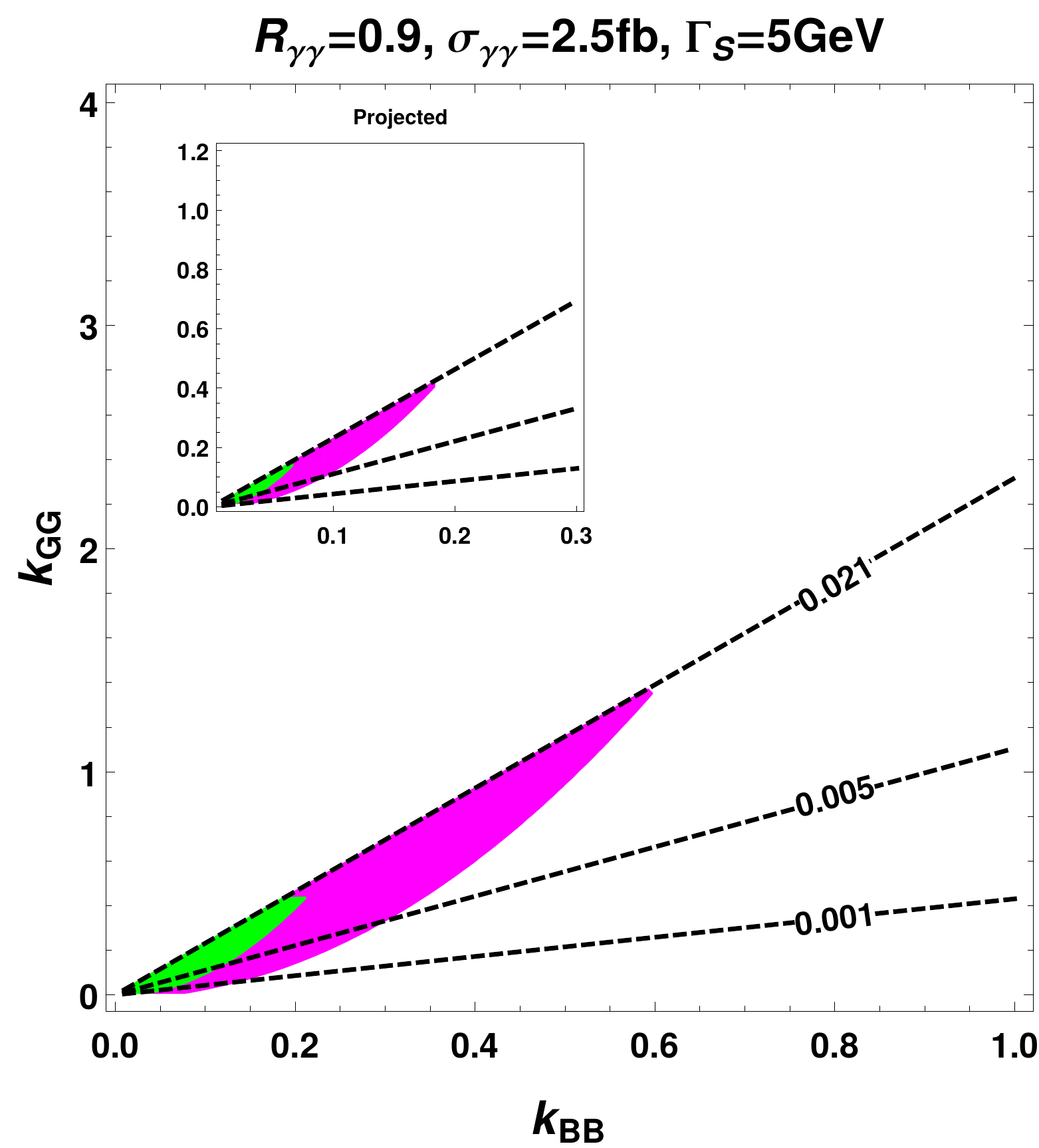}\\
\includegraphics[scale=0.5]{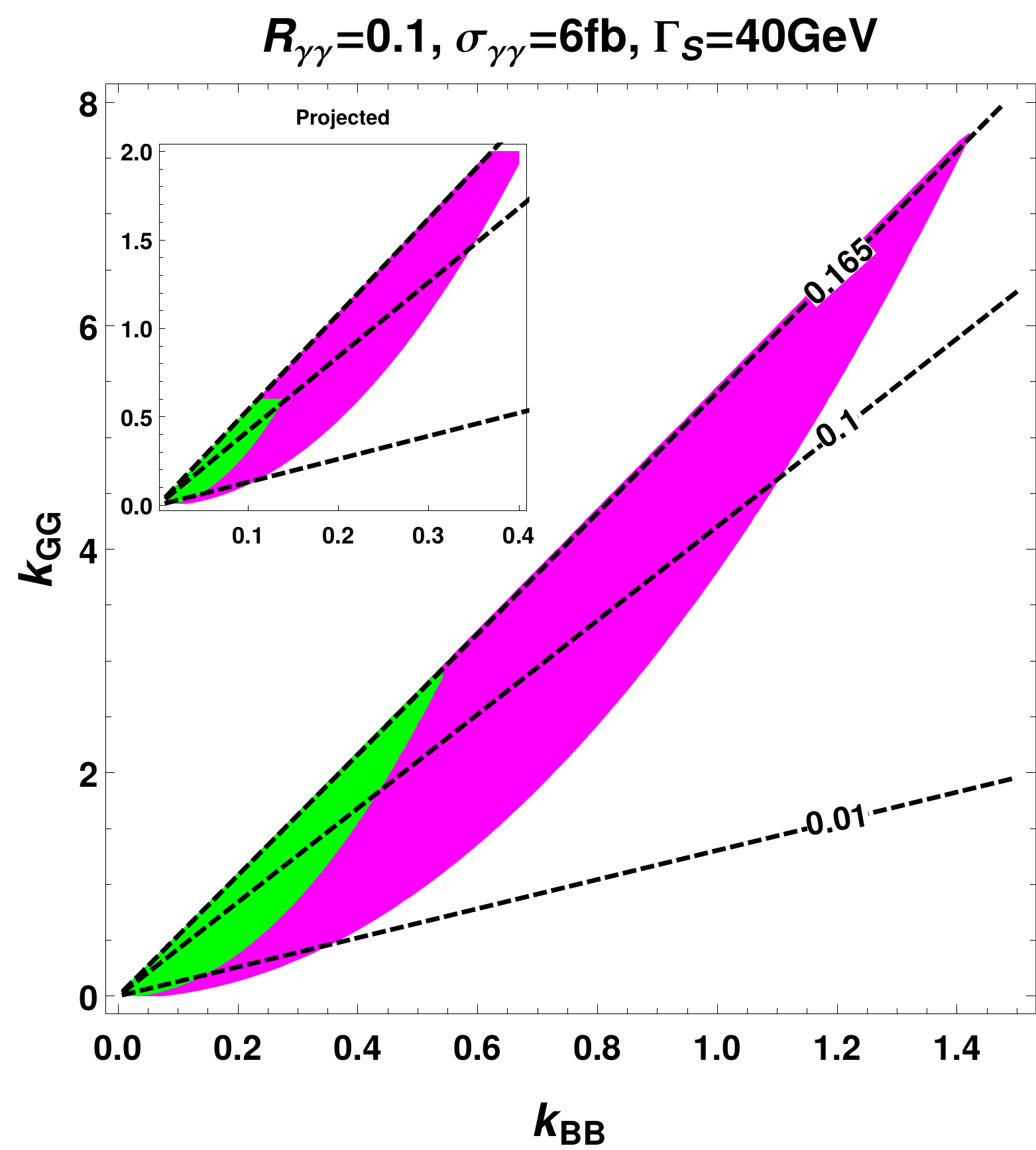}
\includegraphics[scale=0.5]{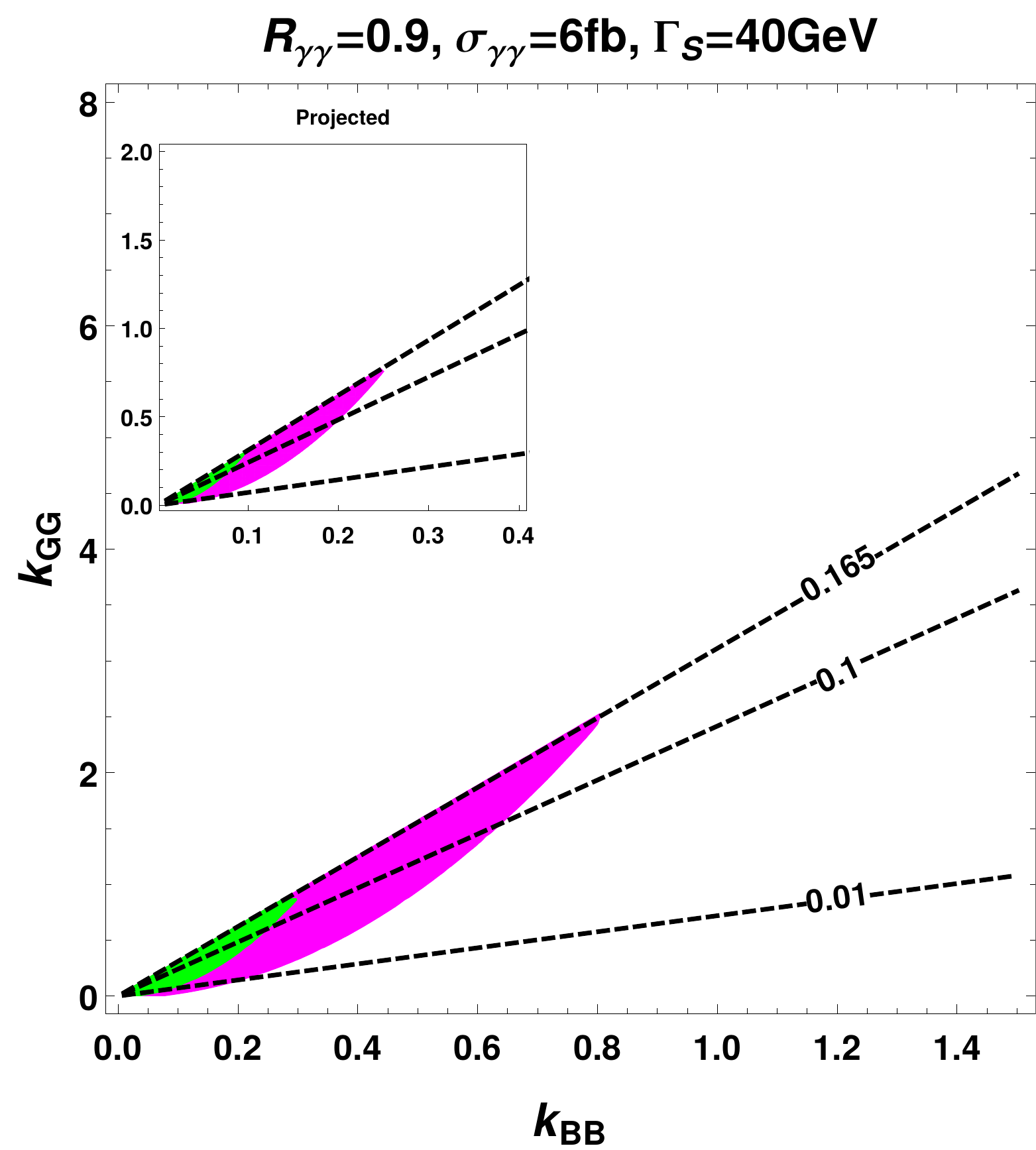}
\caption{The allowed regions in the $k_{BB}$ vs. $k_{GG}$ plane when ALPs are allowed to decay into hadrons. The union of the magenta and green regions represent the portion of the parameters space allowed by the CDF and the ATLAS constraints, ALPs decaying inside the calorimeter, and a total width not larger than the fitted experimental values. The dashed lines have $c_{GG}\frac{\lambda}{\Lambda}$ fixed in order to fit the diphoton cross section and width. The inset plots show the same allowed regions for ten times stronger limits in a hypothetical future search for $Z\to\gamma\gamma(\gamma)$.}
\label{constraints}
\end{figure}

In Figure~(\ref{constraints}) we display the allowed regions in the $k_{GG}$ vs. $k_{BB}$ plane for the wide and narrow scenarios and for small ($R_{\gamma\gamma}=0.1$) and large ($R_{\gamma\gamma}=0.9$) ALP contributions to the diphoton rate. The magenta shaded region represents the allowed region for ALP masses where the CDF bound is relevant, for $0.4<m_a<0.46$ GeV. The green area shows regions escaping the ATLAS bound for $0.46<m_a<3.8$ GeV. The dashed lines have $c_{GG}\frac{\lambda}{\Lambda}$ fixed in order to fit the diphoton cross section and width. Larger $R_{\gamma\gamma}$ means larger contribution from the axionic four-photon channel which requires smaller $k_{BB}$ couplings to fit the diphoton signal.

The first important point is that there is a maximum $c_{GG}\frac{\lambda}{\Lambda}$ compatible with the diphoton signal and still allowed by the experimental constraints and the upper limit on $\Gamma$. In the narrow scenario (the upper plots) this maximum product is $0.021$. As $c_{GG}\frac{\lambda}{\Lambda}$ decreases, the region of allowed points shrinks. For example, for $c_{GG}\frac{\lambda}{\Lambda}=0.001$, only a small intersection for $k_{BB}<0.1$ and $k_{GG}<0.1$ survives if $R_{\gamma\gamma}=0.1$ (the upper left panel). When the ALP contribution to the photonic decay of the 750 GeV scalar is large, the bounds are tighter and the intersections of the dashed lines with the allowed regions are even smaller, as is evident in the upper right panel. In the large width cases displayed in the lower plots of Figure~(\ref{constraints}), the allowed regions are larger than those in the narrow width cases and the maximum $c_{GG}\frac{\lambda}{\Lambda}$ is around $0.17$.

The inset plots show the projected allowed regions for bounds ten times stronger than the present ones. An order of magnitude decrease in the attainable limits at the LHC should be feasible in the near future. Couplings of the ALP with gauge bosons of order 0.1 and smaller could probed in all scenarios from small to large $R_{\gamma\gamma}$ with either $Z\to\gamma\gamma$ or $Z\to\gamma\gamma\gamma$ at the LHC.

\section{A complex scalar $\Phi$} \label{complexscalar}

Having discussed in detail the constraints on the EFT, we now apply these ideas to a concrete model in which $S$ and $a$ belong to the same complex scalar field $\Phi=\frac{1}{\sqrt{2}}(S+v_S+ia)$ which is a singlet under the SM gauge group. We consider a scenario in which this field has a renormalizable  potential invariant under a discrete symmetry $\Phi\rightarrow -\Phi$, but breaking explicitly an $U(1)_a$ symmetry only through quadratic terms $\delta m^2[\Phi^2+(\Phi^*)^2]$. This potential was proposed recently in~\cite{Arcadi:2016dbl} as part of a model of the 750 GeV resonance which also communicates with a dark sector. We also assumed that the interaction between $\Phi$ and the SM is negligible, implying that there is no significant mixing of $S$ with the SM Higgs boson, so the potential is
\bea
V(\Phi)= \mu^2 |\Phi|^2+\lambda_\Phi |\Phi|^4-\delta m^2[\Phi^2+(\Phi^*)^2].
\label{vpot}
\eea
Taking $\mu^2-2\delta m^2<0$, $\lambda_\Phi>0$, the minimum of this potential leads to a vacuum expectation value $\langle\Phi\rangle=v_S/\sqrt2\not=0$. As a result, $S$ and $a$ get masses $m_S=\sqrt{2\lambda_\Phi}v_S=750$ GeV and $m_a=2\delta m$, respectively. The assumption that $m_a\ll m_S$ is a natural one in the sense that in the limit  $\delta m\rightarrow 0$  increase the number of symmetries of the theory -- the  $U(1)_a$ symmetry turns out to be exact having $a$ as its Nambu-Goldstone boson.   

 In order to make the results of our analysis as general as possible, we do not specify the interactions  of $\Phi$ with extra particles but just take into account that, after $\Phi$ get a vacuum expectation value, the effective interactions of $S$ and $a$ with gauge bosons read as follows 
\bea
{\cal L} &\supset & \frac{k_{BB}}{\Lambda}S B^{\mu\nu}B_{\mu\nu}+\frac{k_{WW}}{\Lambda}S W_i^{\mu\nu}W^i_{\mu\nu}+\frac{k_{GG}}{\Lambda}S G_a^{\mu\nu}G^a_{\mu\nu}\nonumber\\
         &+& \frac{k_{BB}}{\Lambda}a\, B^{\mu\nu}\tilde{B}_{\mu\nu}+\frac{k_{WW}}{\Lambda}a\, W_i^{\mu\nu}\tilde{W}^i_{\mu\nu}+\frac{k_{GG}}{\Lambda}a\, G_b^{\mu\nu}\tilde{G}^b_{\mu\nu}.
\label{phiefl}
\eea 
\begin{figure}[!t]
\includegraphics[scale=0.50]{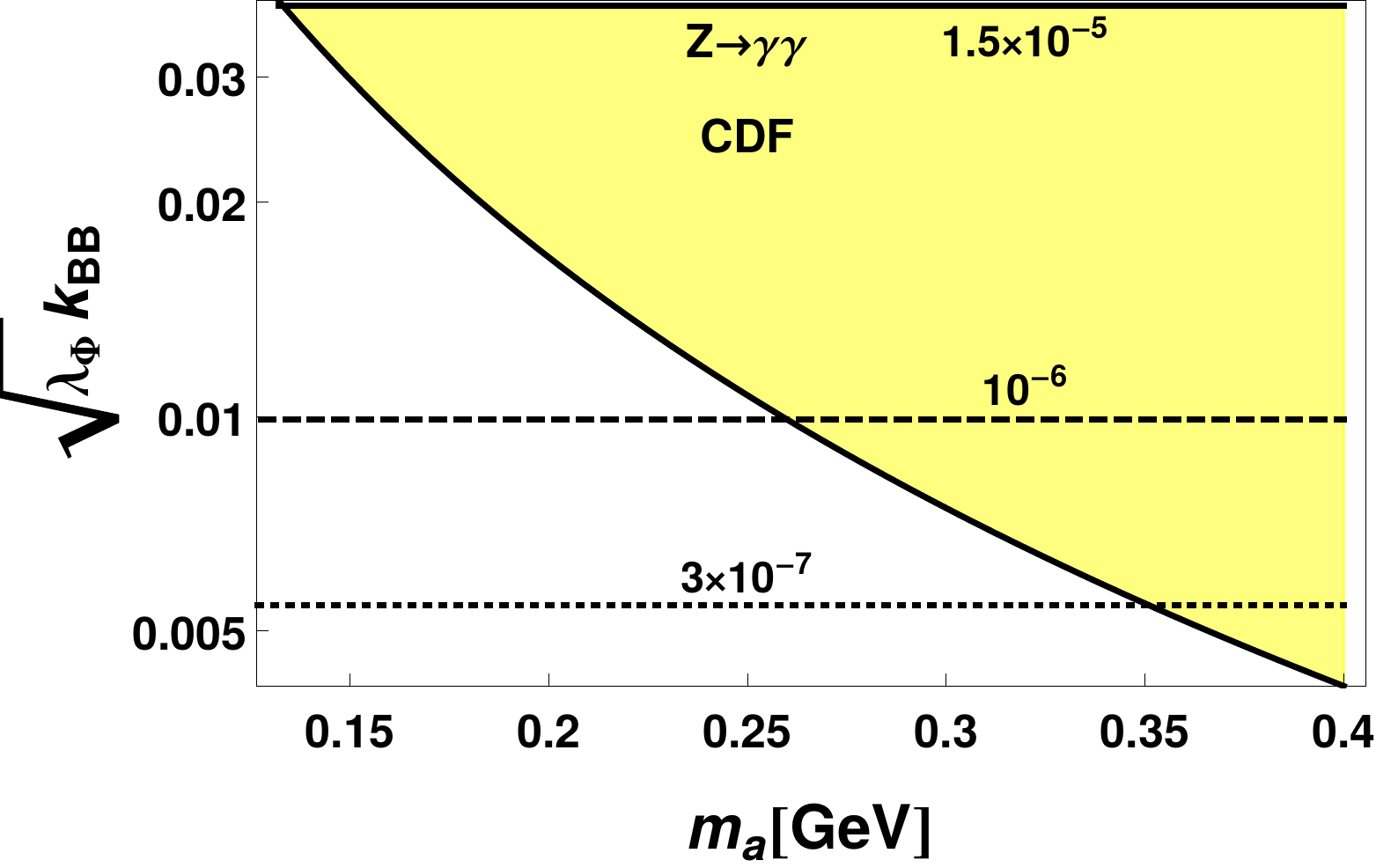}
\includegraphics[scale=0.45]{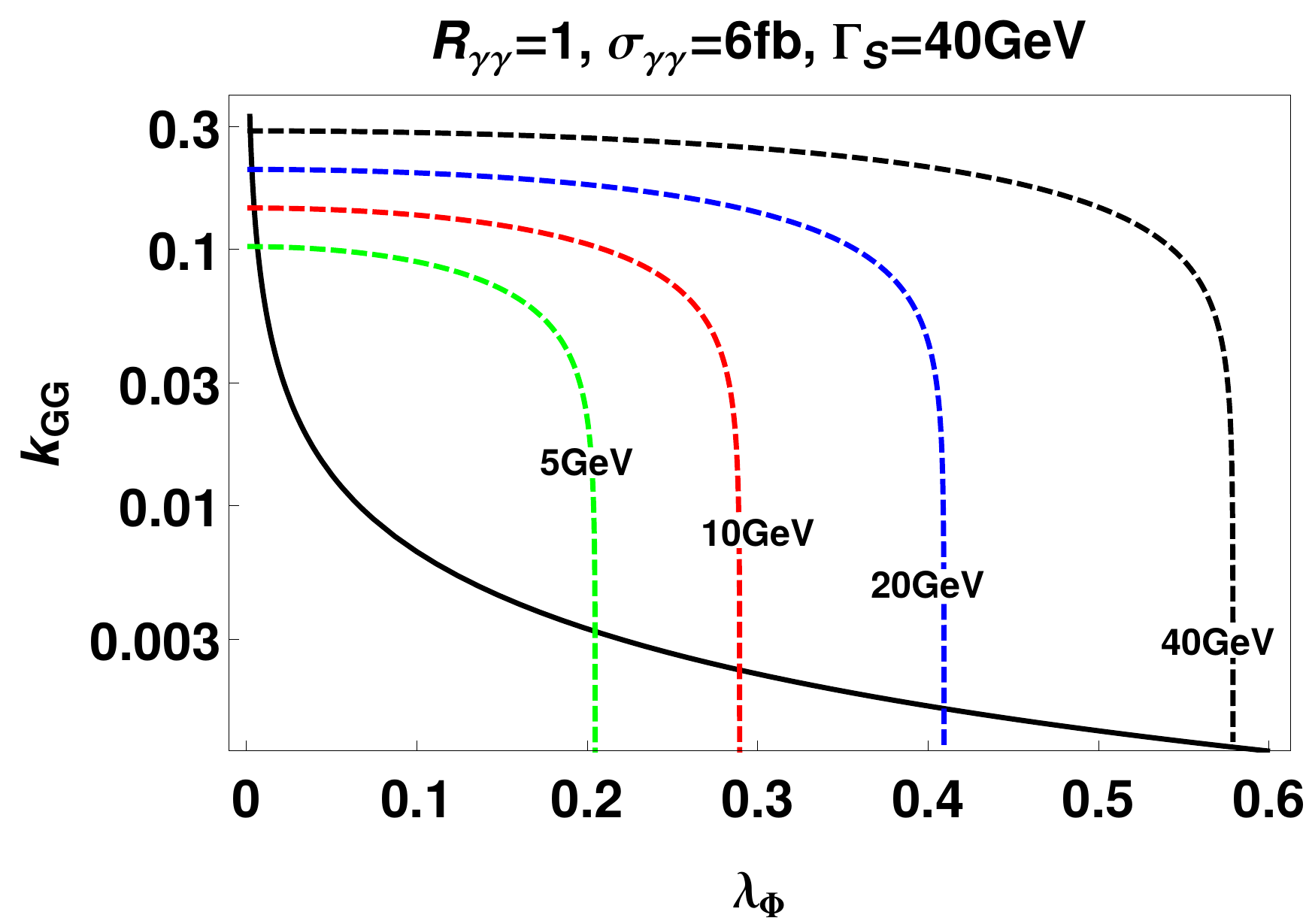}
\caption{$S$ and $a$ belong to the same complex field: The left panel is the analog of Figure~(\ref{fig1}). The yellow shaded region is, as before, the region allowed by the constraints, and the dashed and dotted lines, the projected limits from a future experiment. The right panel is the analog of Figure~(\ref{xsecs}). In these plots we assume $R_{\gamma\gamma}=1$ in the large width scenario.}
\label{yan1}
\end{figure}

In terms of the EFT parameters, the following associations can be made after symmetry breaking
\be
c_{VV}=k_{VV}\to \sqrt{2\lambda_\Phi}\frac{\Lambda}{m_S}k_{VV},\;\; \lambda=\sqrt{\frac{\lambda_\Phi}{2}}m_S
\ee
with $V=G,B,W$. In this case, $S$ and $a$ have the same couplings to  gauge bosons, that is, $k_{BB}=c_{BB}$, $k_{WW}=c_{WW}$ and $k_{GG}=c_{GG}$. For simplicity, we again consider that the particles which are the mediators involved in the loop process generating the the effective interactions in Eq.~(\ref{phiefl}) are singlets under $SU(2)_L$ so that $c_{WW}=k_{WW}=0$. We then have four parameters involved in the analysis: $k_{GG},\; k_{BB},\; \lambda_\Phi$ and $m_a$.

If $m_a<3m_{\pi^0}$ we are able to constrain the product $\lambda_\Phi k_{BB}^2$ from Eqs.~(\ref{bound:ld}--\ref{bound:brz}) as follows
\be
\boxed{\frac{2\pi m_S^3}{10^{16}\times c_w^4m_a^4\times \ell_{decay}}<\lambda_\Phi k_{BB}^2<\frac{3\pi m_S^2\Gamma_Z}{m_Z^3c_w^2s_w^2}}
\ee

Once we have fixed $\lambda_\Phi$ and $k_{BB}$ in the allowed region of the parameter space, $k_{GG}$ can be computed by requiring that
\be
\lambda_\Phi k_{GG}=\sqrt{4\pi^2 R_{\gamma\gamma}C\times\left[\frac{\sigma_{\gamma\gamma}^{13TeV}}{8\hbox{fb}}\right]\times\left[\frac{\Gamma_S}{m_S}\right]}
\label{bounds:yan}
\ee

Let us present now the results for the case where $S$ and $a$ belong to the same complex field. We take $R_{\gamma\gamma}=1$ as in Ref.~\cite{Arcadi:2016dbl}, i.e., the ALP contribution accounts for the entire diphoton signal. In Figure~(\ref{yan1}) we show, in the left panel, the 95\% CL allowed region in the $\sqrt{\lambda_\Phi}k_{BB}$ vs.  $ m_a$ plane in the wide scenario and, in the right panel, the points on the $k_{GG} $ vs. $ \lambda_\Phi$ plane which explain the diphoton rate for fixed widths $\Gamma$.

\begin{figure}[!t]
\includegraphics[scale=0.6]{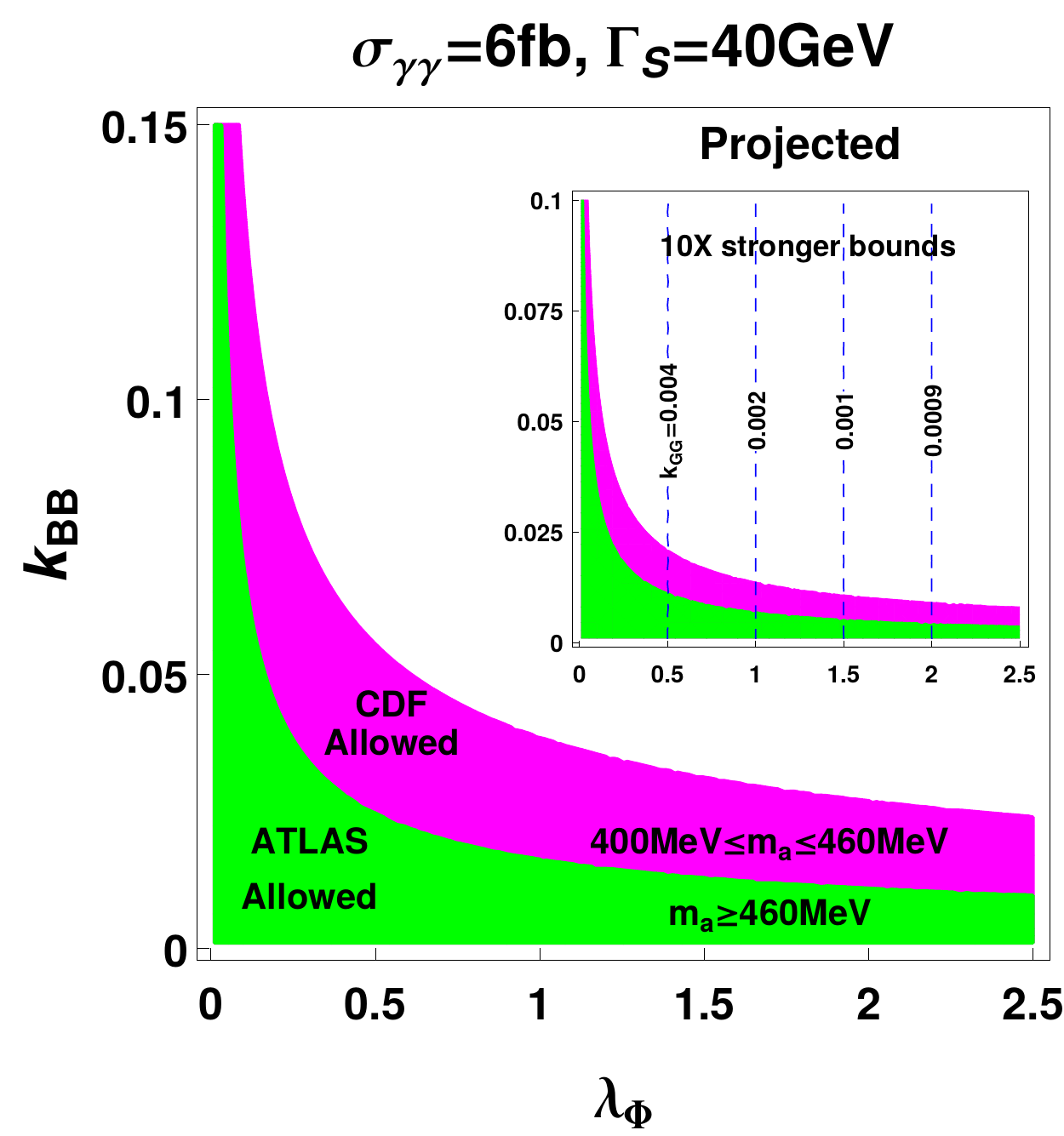}
\caption{$S$ and $a$ belong to the same complex field: The allowed regions of the $k_{BB}$ vs. $\lambda_\Phi$ plane in the large width scenario with $R_{\gamma\gamma}=1$. The interpretations of the shaded areas and the inset plot are the same as those of Figure~(\ref{constraints}) for the EFT parametrization. The vertical dahsed lines in the inset plot give the $k_{GG}$ coupling, as compued from Eq.~(\ref{bounds:yan}), which is necessary to fit the diphoton signal.}
\label{yan2}
\end{figure}

If decays to gluons are allowed, $k_{GG}$ plays a role by decreasing $Br(a\to\gamma\gamma)$ but, once we have fixed $k_{BB}$ and $\lambda_\Phi$ in order to satisfy the bounds of Eq.~(\ref{bounds:yan}), $k_{GG}$ also gets fixed by imposing the fitting of the diphoton signal from Eq.~(\ref{diphoton}) with $c_{GG}=k_{GG}$. The resulting allowed regions in the large width scenario are shown in Figure~(\ref{yan2}). For $0.4< m_a <0.46$ GeV, the union of the green and magenta areas contain points of the parameters space where the ALP decays inside the ECAL, fit the diphoton signal, and are not excluded by the limits on the photonic decays of the $Z$ boson. The smaller green area represents the region allowed by the ATLAS bound on $Z$ decays and $m_a > 0.46 GeV$. The inset plot shows the same allowed regions but for  ten times stronger bounds in a projected future search for $Z\to \gamma\gamma(\gamma)$. The dashed vertical lines show the values assumed by the coupling $k_{GG}$ in those points of the parameters space.
\section{Searching for a $\gamma\gamma$ resonance at the $Z$-pole} \label{Zpoleresults}

\subsection{Cutting and counting}

From the previous Sections, it is clear that if the ALP-$S$-cion connection is correct, we should also see a $\gamma\gamma$ resonance at the $Z$-pole coming from $Z \to a \gamma$. After first discussing the range of ALP masses which are most amenable to a preliminary search, we give details of our simulations and results.

Referring to Fig.~\ref{masses}, we can see that ALPs with masses below 4 GeV but above around 0.46 GeV lead to diphoton signals from the $S$ decay and to three photons signal from $Z$ decays. ALPs lighter than 0.46 GeV can mimic diphoton signals from $S$ as well as $Z\to \gamma\gamma$, since the photons from their decays get too collimated to be resolved in the LHC detectors. Generally, for the entire range of masses between 0.4 GeV and 4 GeV, the two softest photons are very collimated and a dedicated experimental study involving an accurate estimate of cut and identification efficiencies, and photon isolation, should be performed aimed to determine the conditions of operating near the angular detector resolutions. 

To simplify matters somewhat, we restrict ourselves to the simulation of $Z$ decaying to two photons only, that is, where the ALP decays into photons pairs so collimated that they cannot be resolved in the LHC detectors. We will therefore take benchmark ALP masses below around 0.4 GeV, and assume what we believe are reasonable values for the fake jet rate and the identification efficiency of the photon-jet.

We simulated parton level events for signal $pp\to Z\to a\gamma\to \gamma\gamma\gamma$, and the SM backgrounds of continuum production of $\gamma\gamma$, $j\gamma$, $jj$, $\gamma\gamma\gamma$ at Leading Order with \texttt{MadGraph5}~\cite{Alwall:2014hca,Alwall:2014bza}. After that we simulated parton hadronization and showering with \texttt{Pythia6}~\cite{Sjostrand:2006za} and detector effects with \texttt{Delphes3}~\cite{deFavereau:2013fsa}. The photon isolation criteria was based on the ATLAS experimental study of Ref.~\cite{Aad:2015bua}-- rejecting events where particles fall within a cone of radius 0.15 around the candidate photon with a deposit of energy larger than 4 GeV.
\begin{figure}[!t]
\includegraphics[scale=0.5]{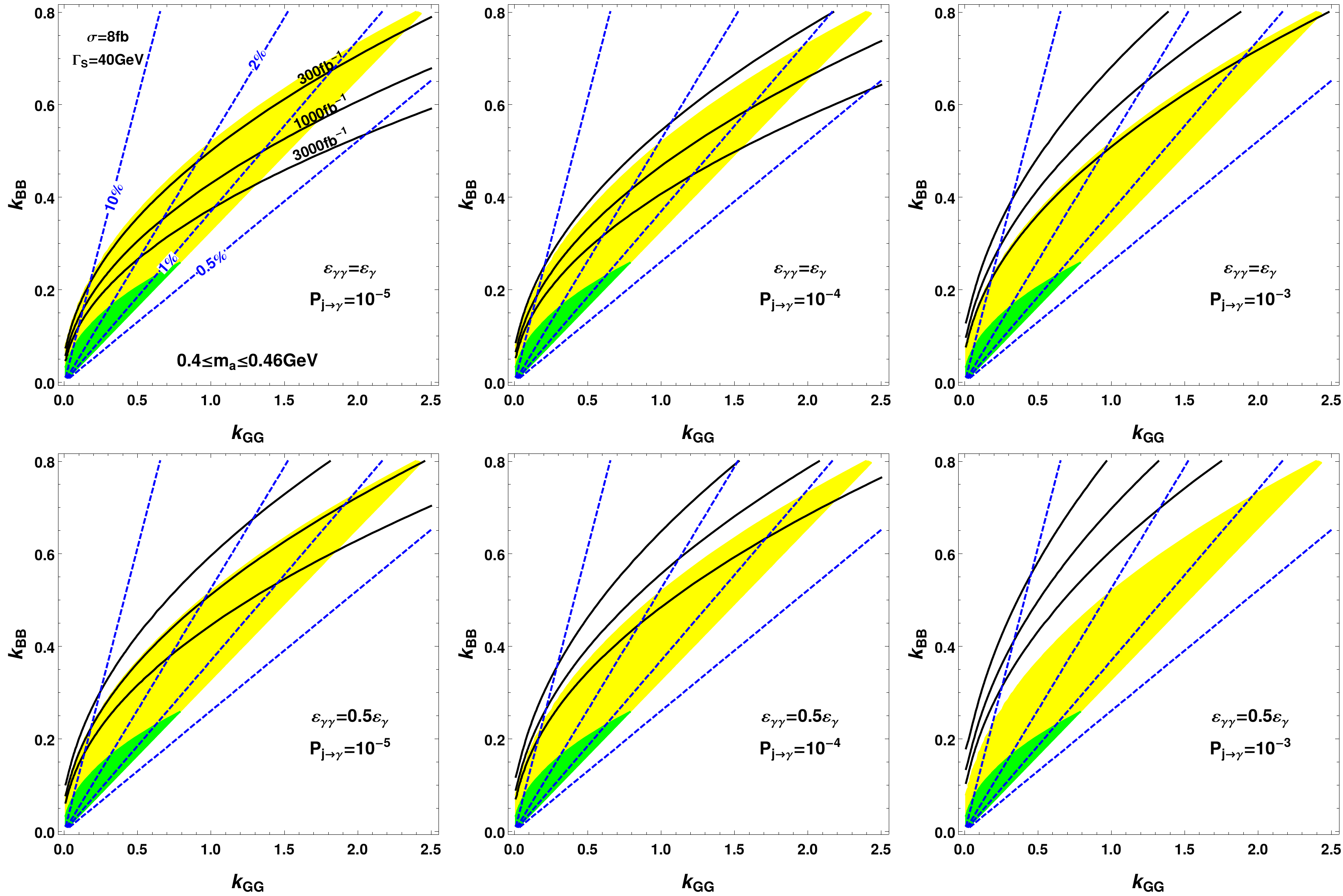}
\caption{In all these plots, the union of the yellow and green areas are the points satisfying the constraint from the CDF Collaboration and a total width not larger than the fitted value in the wide scenario. The green area is the allowed region for a ten times stronger limit. The solid lines are contour lines where a $5\sigma$ discovery is possible with 300, 1000, and 3000 fb$^{-1}$. Dashed lines have constant branching ratios of ALPs decaying to photons. We assume the same identification efficiency for single photons and {\it photon-jets} in the upper row of plots for three different fake jet rejection factors from $10^{-3}$ to $10^{-5}$. In the lower row the {\it photon-jet} efficiency is fixed at the half of single photon efficiency.}
\label{discovery1}
\end{figure}

The acceptance cuts adopted are given by
\be
p_T(\gamma) > 20\,\hbox{GeV}\;\; ,\;\; |\eta(\gamma)|<2.0
\ee
for both photons.

For ALP masses below 0.4 GeV, the majority of signal events contain indeed just two photons. In order to select photons at the $Z$-pole we imposed the additional cut on the photons invariant mass $m_{\gamma\gamma}$
\be
85\,\hbox{GeV}< m_{\gamma\gamma}< 95\,\hbox{GeV}
\ee

The signal cut efficiency is $\sim 30$\% while the background rejection is at least 0.02. We present in Figure~(\ref{discovery1}), the contour lines in the $k_{GG}\times k_{BB}$ plane where a $5\sigma$ discovery of $Z$ bosons decaying to gamma rays is possible for 300, 1000 and 3000 fb$^{-1}$ of integrated luminosity, $0.4<m_a <0.46$ GeV where $Br(a\to\gamma\gamma)<1$, and $R_{\gamma\gamma}=1$ in the wide scenario. In the yellow and green shaded areas $Br(Z\to \gamma\gamma)<1.5\times 10^{-5}$ respecting the CDF bound and the total width of the scalar $S$ is less than 40 GeV. The green area represents a projected upper bound ten times stronger than the current CDF limit. We immediately see that if the LHC is able to reach an upper limit of order $10^{-6}$, than it will be very hard to discover these $Z$ bosons in the LHC. The dashed blue lines are lines with constant $Br(a\to\gamma\gamma)$. The only effect of decreasing $R_{\gamma\gamma}$ is to enlarge the allowed region as we see in Figure~(\ref{constraints}).

We also found that the signal significance depends strongly on the fake jet rate and the identification efficiency of the photon-jet. The ability to reject jets faking photons is crucial to reduce the $j\gamma$ and $jj$ backgrounds which have 10 and $10^5$ larger cross sections, respectively, compared to the $\gamma\gamma$ background. In the plots we varied the probability of a jet being taken as a photon $P_{j\to\gamma}$ from $10^{-3}$ to $10^{-5}$. The other important tagging factor is the photon efficiency of the photon-jet. The two collimated photons will hit a single calorimeter cell for very light ALPs and will be detected with an efficiency $\varepsilon_{\gamma\gamma}$. In principle, this efficiency could be different from the single photon efficiency and a dedicated experimental study or a very careful simulation of the detectors should be done in order to estimate it. We thus chose to work with an optimistic factor of $\varepsilon_{\gamma\gamma}=\varepsilon_\gamma$, where $\varepsilon_\gamma$ is the single photon efficiency taken as in the \texttt{Delphes3} package, varying between 0.85 and 0.95 depending on the photon's transverse momentum and rapidity, and a pessimistic factor 50\% smaller than the optimistic one.
\begin{figure}[!t]
\includegraphics[scale=0.5]{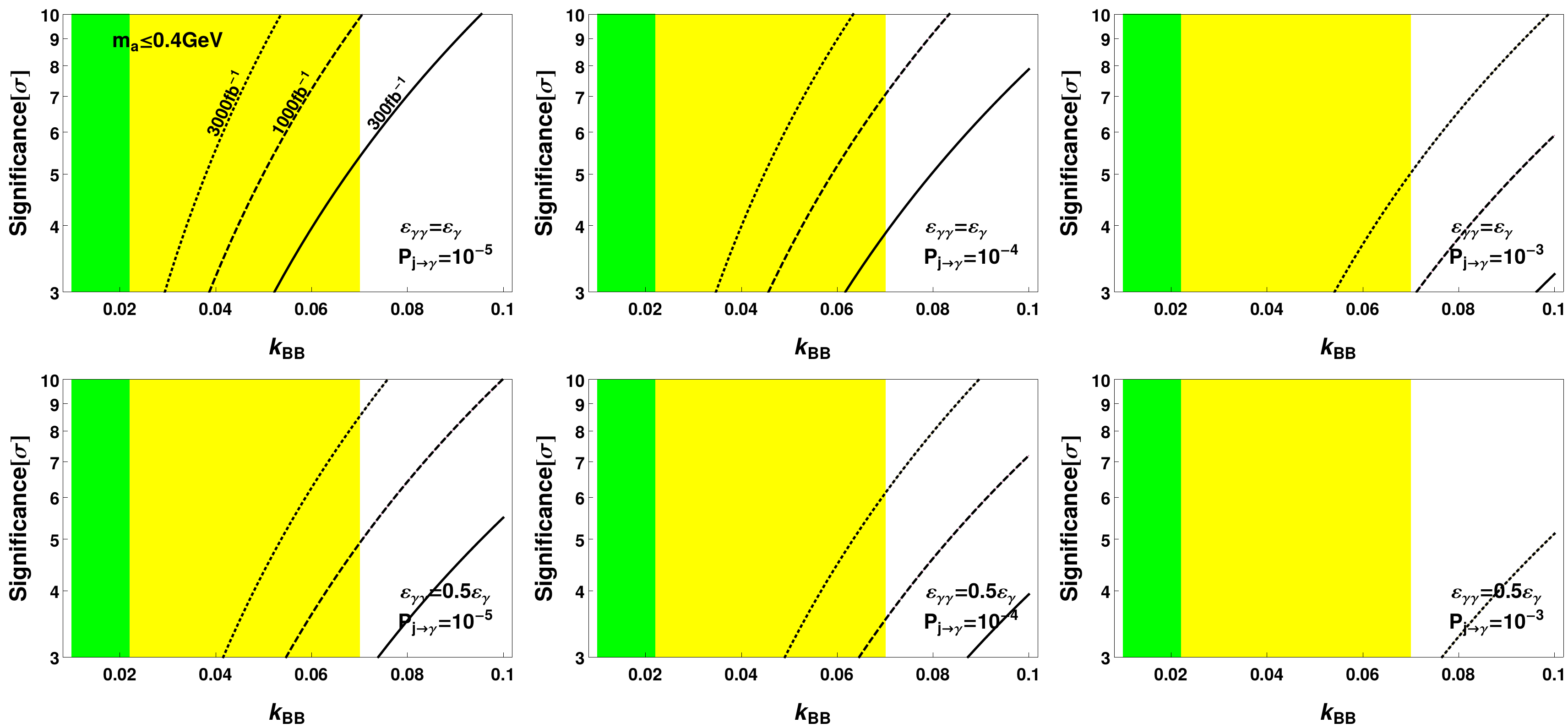}
\caption{The significance of a diphoton signal from $Z$ decays in axion-like models of the 750 GeV resonance for ALPs with 100\% decays to photons. We, again, fix three different luminosities and fake jet rejection factors, and two different {\it photon-jet} efficiencies to illustrate our results as in the previous plot. The meaning of the colors is the same of the previous figure.}
\label{discovery2}
\end{figure}
In the three upper(lower) panels of Figure~(\ref{discovery1}) we are optimistic(pessimistic) about the photon-jet efficiency. If events with jets faking jets could be rejected at the $10^{-5}$ rate, we see in the left upper plot that it is possible to discover the photonic decay mode of the $Z$ boson with 300 fb$^{-1}$ if $k_{BB}$ is not too small for a given $k_{GG}$. Of course, as $k_{BB}$ drops more luminosity is needed. The discovery becomes increasingly hard as $P_{j\to\gamma}$ increases. On the other hand, for $\varepsilon_{\gamma\gamma}=0.5\varepsilon_\gamma$, discovery will only be possible with around 1 ab$^{-1}$ if $p_{j\to\gamma}=10^{-5}$, and no discovery at all will be possible if $p_{j\to\gamma}=10^{-3}$. The plots for the narrow scenario are identical to these ones, but in a smaller region extending itself up to 0.6(1.4) in the $k_{BB}$($k_{GG}$) direction.

Now, if the ALP is lighter than 0.4 GeV, $Br(a\to\gamma\gamma)=100$\% and the discovery becomes possible in the more optimistic scenario shown in the left upper panel of in Figure~(\ref{discovery2}). In this case, with 300 fb$^{-1}$, $k_{BB}$ couplings of 0.07 can be probed and up to 0.03 for 3 ab$^{-1}$. Again, we observe that if the LHC collaborations are able to exclude branching fractions of $Z$ bosons decaying to photons pairs of order $10^{-6}$, it will be very difficult to discover this signal in this cut-and-count approach.

\subsection{Discovery estimate from an improved analysis for detecting photon-jets}

Identifying {\it photon-jets} at hadron colliders has been an interesting line of investigation in recent years~\cite{Dasgupta:2016wxw,Draper:2012xt,Ellis:2012sd,Ellis:2012zp}. In particular, concerning the diphoton excess of 750 GeV photons pairs, in Ref.~\cite{Dasgupta:2016wxw} the authors show that is possible to discern between axion-like models where the $S$-cion decays to two pairs of pairs of collimated photons through the interaction of a light pseudoscalar within an effective model very similar to what we are considering here. The idea is basically counting the number of photons conversions to $e^+e^-$ pairs in the inner detector. In the ATLAS, for example, four out of ten photons are expected to be converted into electron-positron pairs and the ability to recognize such pairs as coming from photons produced at the interaction point is very important to reach a high photon identification rate. More photons hitting a given cell means more $e^+e^-$ conversions, so from this basic fact it is possible to tell if more photons than the expected from single isolated photons are being converted and, then, evaluate the likelihood of a model compared to some different hypothesis. In Ref.~\cite{Dasgupta:2016wxw}, an axion-like model of the 750 GeV resonance can be distinguished from models where $S$ decays promptly to two isolated photons at the statistical level of 2(5)$\sigma$ with $\sim 30$(100) events.

Of course, the very same technique could be used to increase the discerning power of identification for photon-jets from $Z$-decays. But not only this. As remarked in Ref.~\cite{Dasgupta:2016wxw}, there are other ways to tell if the detector was hit by a single isolated photon or a photon-jet, for example, by choosing appropriated photon isolation criteria or observing that in events where just one of the two collimated photons is converted, the ratio $(p_T\; \hbox{of the track})/E_{CAL}$ is not the expected from a single converted photon track.

A dedicated analysis in identifying photon-jets in environments rich in QCD jets and isolated photons was performed in Refs.~\cite{Ellis:2012sd,Ellis:2012zp}. In these works, samples of Higgs bosons of 120 GeV were assumed to decay to light scalars which, by their turn, decay to collimated photons. Using substructure techniques and by training decision trees to an efficient separation of signal and backgrounds events, a fake-QCD jet rate of order $10^{-4}$--$10^{-5}$ and a fake-single photon rate of order $10^{-1}$--$10^{-4}$ can be obtained depending on the mass of the light scalar. The less efficient separation rates occur for the lighter scalars.

These misidentification tagging rates can be used for a rough estimate of how much we expect that this kind of dedicated analysis facilitates the discovery of signals with photons at the $Z$-pole as suggested in this work. As $Z$ bosons and 120 GeV Higgs bosons have more or less similar masses, we expect that the kinematics of the photons are not too different in both case. For example, as $Z$'s and Higgses decay predominantly in the central region with similar transverse momenta, the $e^+e^-$ probabilities for each kind of event is similar too. Then, based on Refs.~\cite{Ellis:2012sd,Ellis:2012zp}, we fix $0.1$ as the fake-single photon rate and $2\times 10^{-4}$ for the  fake-QCD jet rate, and adjusting the \texttt{Delphes3} photon ID efficiency to 80\%. Cut efficiencies are taken as the cut analysis performed in the later subsection.

\begin{figure}[!t]
\includegraphics[scale=0.5]{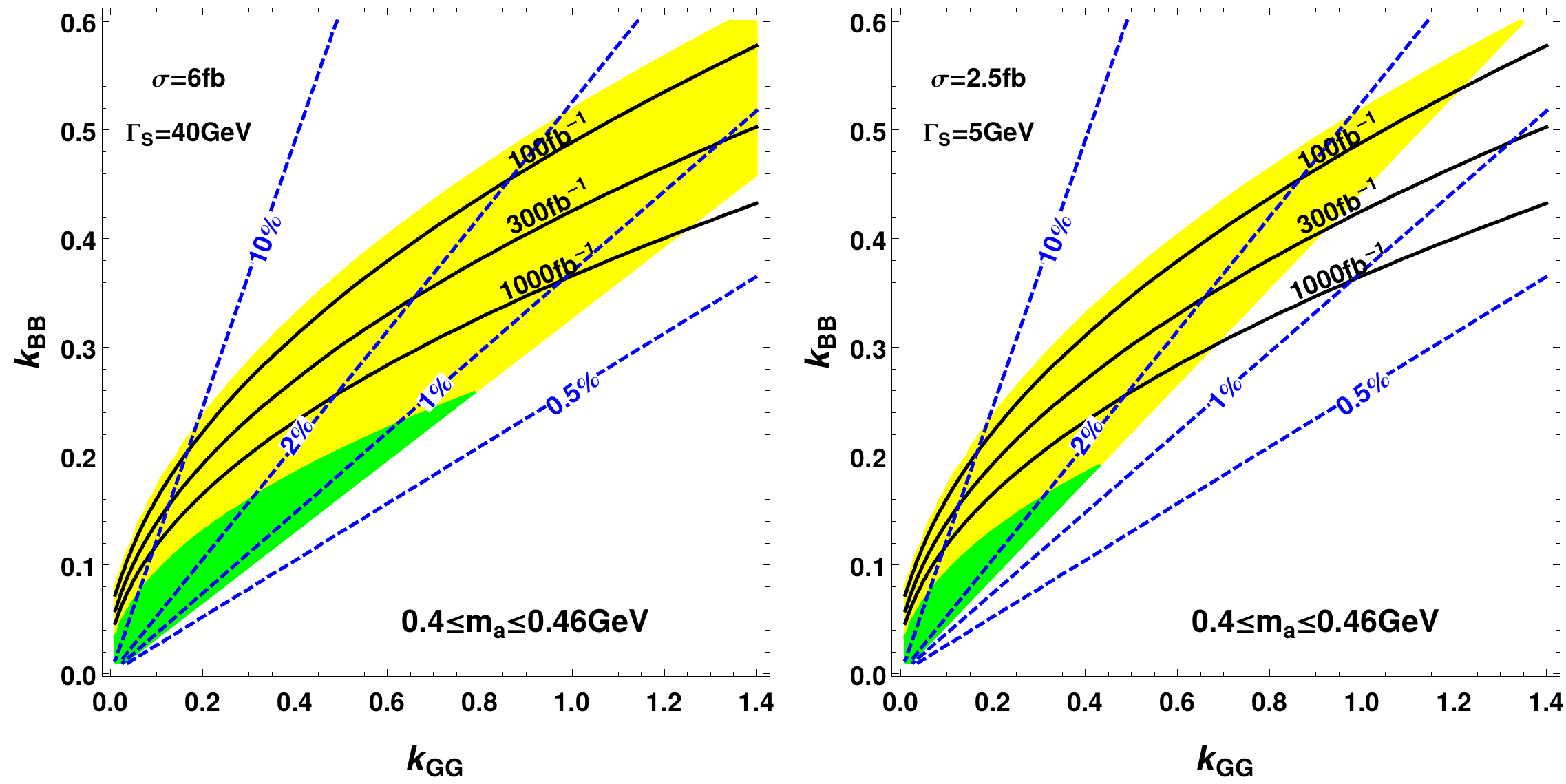}\\
\includegraphics[scale=0.5]{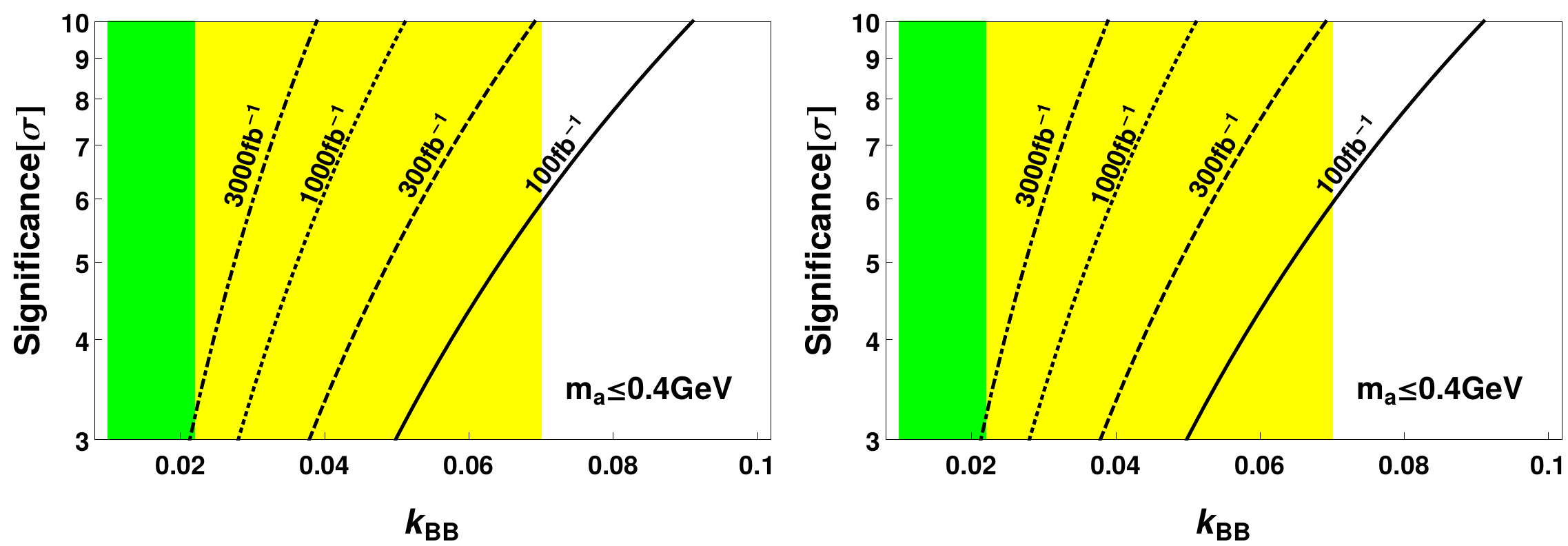}
\caption{In all the upper plots, the union of the yellow and green areas are the points satisfying the constraint from the CDF Collaboration and a total width not larger than the fitted value in the wide(narrow) scenario at the left(right) plot. The green area is the allowed region for a ten times stronger limit. The solid lines are contour lines where a $5\sigma$ discovery is possible with 100, 300, and 1000 fb$^{-1}$. Dashed lines have constant branching ratios of ALPs decaying to photons. The identification efficiency and fake-QCD jet and fake-single photons rates are given in the text. In the lower plots, the yellow and green areas have the same meaning, but for the case of a complex scalar interacting according to Eq.~(\ref{vpot},\ref{phiefl}) and $m_a\leq 0.4$ GeV.}
\label{discovery3}
\end{figure}

We show in Fig.~(\ref{discovery3}), the estimate of the discovery reach of the 13 TeV LHC to observe $Z\to\gamma\gamma$ at $5\sigma$ in the upper plots assuming the EFT model of Eq.~(\ref{eft}), and the singlet scalar complex of Eq.~(\ref{vpot},\ref{phiefl}) in the two lower plots. We immediately see that this dedicated analysis with photon-jets has a greater potential to enlarge the region of the parameters space for discovery.

Contrary to the cut-and-count analysis, improving the tagging efficiency for photon-jets allows us probe the currently permitted parameter space (the yellow shaded areas) with 100 fb$^{-1}$ as we see in the upper plots of Fig.~(\ref{discovery3}) in the narrow and the wide scenarios for the EFT model. For the concrete model presented in section~(\ref{complexscalar}), $k_{BB}$ couplings of around 0.5 can be probed with 300 fb$^{-1}$ compared to 0.7 of the previous analysis as we see in the lower plots of Fig.~(\ref{discovery3}).

It is beyond the scope of this investigation to go much further into this direction, but we believe that a dedicated study along the lines of Ref.~\cite{Dasgupta:2016wxw,Ellis:2012sd,Ellis:2012zp} could boost the discovery prospects for photons at the $Z$-pole at the LHC, be it related or not to the 750 GeV $S$. Essential ingredients for this study are an accurate estimate of the photons rapidity and transverse momentum distributions, once the probability of $e^+e^-$ conversion depends on these kinematic features of the event, and good discriminants against the $Z\to e^+e^- +\gamma$ background, with a bremsthralung $\gamma$.


\section{Conclusions}

New physics beyond the Standard Model may be right around the corner with the emergence of tantalizing signals of a new resonance in the $\gamma\gamma$ channel with mass around 750 GeV in both ATLAS and CMS experiments in the LHC. If a new scalar is confirmed in this run of $pp$ collisions, a whole new dynasty of fundamental particles might be revealing themselves. 

In this work, we have considered scenarios in which the 750 GeV resonance decays to ALPs, which further decay to photons and mimic the signal. These models are motivated by several factors. They come naturally in Hidden Valley scenarios, are able to accommodate a large width of $S$ within the perturbative regime, and open up connections to models of cold dark matter.

As in the beginnings of the Standard Model, we are still ignorant about the interactions responsible for the production of the scalar $S$ and its decays to electroweak gauge bosons.  An EFT parametrization of couplings of $S$ and $a$ to capture loop-level interactions allows one to constrain them. The interaction between $S$ and $a$ is assumed to be renormalizable at the tree level and the decay channel $S\to aa$ can be responsible for the large  width of $S$. In order to fit the observed diphoton excess and evade the collider constraints from searches of resonances in the $ZZ$, $WW$, $Z\gamma$ and $gg$ channels more easily, an ALP $a$ lighter than a few GeV contributes to the diphoton signals through its own decays to very collimated diphotons. If the mass of the ALP is small enough, those diphotons are emitted so colinearly that they hit the same cell of the electromagnetic calorimeter of the detectors mimicking a single hit. In this way, the process $pp\to S\to aa\to \gamma\gamma\gamma\gamma$ will lead effectively to events with diphotons.

It is hard to think of UV completions, however, where the ALP couples only to photons and not to the other electroweak gauge bosons. Gauge invariance would seem to dictate an interaction of ALPs with at least the $Z$ boson, in the case the ALP is an $SU(2)_L$ singlet. A light ALP leading to collimated photons cannot decay to electroweak bosons, but a $Z$ boson can decay as $Z\to a\gamma\to \gamma\gamma\gamma$. It turns out that in axion-like models of the 750 GeV resonance, the branching ratio of the photonic decays of the $Z$ boson is orders of magnitude larger than what is expected in the Standard Model ($\sim$ 10$^{-10}$) and can be so large that collider constraints from the search of $Z$ bosons decaying to photons need to be taken into account.

In this work, we show that the parameters of an EFT description of the 750 GeV signal get bounded by experimental limits on the two and three photons decays of the $Z$ boson from the Tevatron and the LHC. We also take into account that (i) the ALP should decay into two collimated photons inside the electromagnetic calorimeter and (ii) the total width of $S$ is bounded by the current experimental best fit value. We consider several scenarios assuming narrow and wide resonances, ALPs decaying exclusively to photons when their masses are smaller than three pion masses and heavier ALPs with additional gluonic decays, and dominant and non-dominant $S\to 4\gamma$ decays compared to the $S\to \gamma\gamma$ decays. For example, if $Br(a\to\gamma\gamma)=1$, limits from the CDF Collaboration on $Z\to\gamma\gamma$ impose an upper bound on the effective ALP-photon coupling of 0.07. If the LHC pushes this limit to a level 50 times stronger, almost all the parameter space of these models can be excluded at 95\% CL. We also show how a concrete model where $S$ and $a$ are the real and the imaginary parts of a complex scalar, respectively, has its parameters bounded by these experimental constraints.

The $Z\to a\gamma\to\gamma\gamma\gamma$ is a striking prediction of these kinds of models. It would be natural, then, to look for photonic decays of the $Z$ boson in the 13 TeV LHC. We estimate the prospects to discover this decay mode of the $Z$ assuming several photon detection efficiencies for both single isolated and bunches of collimated photons. A simple cut-and-count analysis suffices to probe the allowed parameter space of the EFT models with 300 fb$^{-1}$ with optimal photon detection efficiencies. We also provide a simple estimate based on the machine learning analysis of Ref.~\cite{Ellis:2012sd} to better recognize events with photon-jets and find that with 100 fb$^{-1}$ of accumulated data, observing photonic $Z$ decays of models fitting the 750 GeV resonance is possible.

\section{Acknowledgments}

We would like to thank Andrew Askew, Teruki Kamon, Yann Mambrini, Paul Padley, and David Toback. A. Alves and A.G. Dias acknowledge financial support from the Brazilian agencies CNPq, under the processes 303094/2013-3
(A.G.D.), 307098/2014-1 (A.A.), and FAPESP, under the process 2013/22079-8 (A.A. and A.G.D.).

\end{document}